\documentclass[aps,pra,twocolumn,showpacs,nofootinbib]{revtex4}

\usepackage[english]{babel}
\usepackage{amsmath}
\usepackage{amsthm}
\usepackage{amssymb}
\usepackage{booktabs}
\usepackage{color}
\usepackage{hyperref}
\usepackage{cases}
\hypersetup{
  colorlinks   = true,  %Colours links instead of ugly boxes
  urlcolor     = blue,  %Colour for external hyperlinks
  linkcolor    = blue,  %Colour of internal links
  citecolor    = red    %Colour of citations
}
\usepackage{natbib}
\usepackage{pgfplots}
\usepackage{subfigure}
\usepackage{url}

\newcommand{\refFigure}[1]{{\textrm{Fig.~\ref{#1}}}}
\newcommand{\refTable}[1]{{\textrm{Table~\ref{#1}}}}
\newcommand{\refEquation}[1]{{\textrm{Eq.~(\ref{#1})}}}

\newcommand{\del}[1]{\textcolor{black}{#1}}

\usepackage{scalerel,stackengine}
\stackMath
\newcommand\reallywidehat[1]{%
\savestack{\tmpbox}{\stretchto{%
  \scaleto{%
    \scalerel*[\widthof{\ensuremath{#1}}]{\kern-.6pt\bigwedge\kern-.6pt}%
    {\rule[-\textheight/2]{1ex}{\textheight}}%WIDTH-LIMITED BIG WEDGE
  }{\textheight}% 
}{0.5ex}}%
\stackon[1pt]{#1}{\tmpbox}%
}
\usepackage{comment}
\usepackage{float}

\begin{document}

\title{Scattering hypervolume for ultracold bosons from weak to strong interactions}
%\title{\ins{Collisions of three ultracold bosons: from weak to strong interactions}}
%\title{\ins{Scattering of three free ultracold bosons}}
%\title{Elastic scattering of three ultracold bosons}
%\title{Scattering of three free ultracold bosons}
%\title{\ins{Elastic scattering properties of three ultracold bosons}}

\author{P. M. A. \surname{Mestrom}}
%\email{p.m.a.mestrom@tue.nl}
\altaffiliation[Corresponding author: ]{p.m.a.mestrom@tue.nl}

\author{V. E. \surname{Colussi}}

\author{T. \surname{Secker}}

\author{S. J. J. M. F. \surname{Kokkelmans}}
\affiliation{Eindhoven University of Technology, P.~O.~Box 513, 5600 MB Eindhoven, The Netherlands}

\date{\today}


\begin{abstract}
%Determining the elastic scattering properties of three bosons at low energy is a fundamental problem.
The elastic scattering properties of three bosons at low energy enter the many-body description of ultracold Bose gases via the three-body scattering hypervolume $D$.
We study this quantity for identical bosons that interact via a pairwise finite-range potential.
Our calculations cover the regime from strongly repulsive potentials towards attractive potentials supporting multiple two-body bound states and are consistent with the few existing predictions for $D$.
We present the first numerical confirmation of the universal predictions for $D$ in the strongly interacting regime, where Efimov physics dominates, for a local nonzero-range potential. 
%Our findings highlight how finite-range effects, such as \ins{$d$-and $g$-wave} interactions, become important as the interaction strength is reduced.
%\del{As the interaction strength is reduced, additional three-body quasibound states with strong $d$-wave and/or $g$-wave characteristics appear at the zero-energy threshold.}
Our findings highlight how $D$ is influenced by three-body quasibound states with strong $d$-wave or $g$-wave characteristics in the weakly interacting regime.
\end{abstract}

\maketitle

\textit{Introduction.}---Due to the precise experimental control of interatomic interactions via external magnetic fields, ultracold atomic gases have emerged as a versatile field for studying and manipulating quantum systems.
The effective two-body interaction strength given by the $s$-wave scattering length $a$ can be tuned via Feshbach resonances \cite{chin2010feshbach}.
When $|a|$ diverges, Efimov predicted the existence of an infinite number of three-body bound states whose universal scaling properties have been observed experimentally \cite{efimov1970energy, efimov1971weakly, braaten2006universality, naidon2017review, dincao2018review,kraemer2006cesium133, huang2014cesium133}. 
This nonperturbative three-body effect influences the properties of strongly interacting Bose gases \cite{colussi2018dynamical, braaten2011contacts, werner2012contacts, colussi2018contact3body, colussi2019contact3body} and Bose-Einstein condensates (BECs) interacting with an impurity particle \cite{levinsen2015bosepolaron, yoshida2018bosepolaron, sun2017bosepolaron, naidon2018impuritiesInBEC}. Connecting few-body processes with bulk properties of ultracold Bose gases is fundamental to our understanding of these quantum many-body systems.

This connection is evident from a low-density expansion of the ground-state energy density $\mathcal{E}$ of a dilute BEC with a homogeneous number density $n$ \cite{tan2008hypervolume}:
\begin{eqnarray}\label{eq:energy_density_BEC}
\mathcal{E} &= \frac{2 \pi \hbar^2 n^2 a}{m} 
\Bigg\{
1 + \frac{128}{15 \sqrt{\pi}} \sqrt{n a^3}
+ \Bigg[ \frac{8 (4 \pi -  3 \sqrt{3})}{3} \text{ln}(n a^3) \nonumber
\\
&
+ \frac{D}{12 \pi a^4} 
+ \pi r_s/a + 118.5 \Bigg] n a^3 + ...
\Bigg\}
\end{eqnarray}
where the dots indicate higher-order correction terms in the diluteness parameter $n a^3$, $m$ is the mass of a boson, $r_s$ is the two-body effective range, and $a>0$.
The $\sqrt{n a^3}$ correction, calculated by Lee, Huang, and Yang (LHY) \cite{lee1957LHYprogressreport, lee1957LHY}, originates from two-body elastic scattering characterized by $a$ alone. 
Experiments have probed LHY physics by measuring the critical temperature of a BEC \cite{smith2011criticalTemperature}, quantum depletion \cite{lopes2017depletion}, excitation spectrum \cite{papp2008excitation, lopes2017quasiparticle}, thermodynamic equation of state \cite{navon2011dynamicsBEC} and contact \cite{wild2012rubidium85}. 
Additionally, recent studies predicted \cite{petrov2015dropletsBoseBoseMix} and experimentally confirmed the formation of quantum droplets in mixtures \cite{cabrera2018quantumdroplet, tarruell2018quantumdroplet, semeghini2018quantumdroplet} and dipolar BECs \cite{kadau2016dipolarquantumdroplet, ferrierbarbut2016dipolarquantumdroplet, chomaz2016dipolarquantumdroplet} due to a stabilizing force originating from the LHY correction.

As the study of strongly interacting Bose gases advances, there is the opportunity to observe beyond-LHY corrections.
These corrections have been studied both phenomenologically, via extensions of the Gross-Pitaevskii equation \cite{akhmediev1999BECwith3body, gammal2000BECwith3body, gammal2000BECwith3bodyPRA, adhikari2002BECwith3body, bulgac2002droplets, bedaque2003droplets, blakie2016dropletsDipolar, al_jibbouri2013BECwith3body, xi2016BECwithg3Droplet},
and microscopically, via quantum Monte Carlo simulations \cite{giorgini1999groundstateBEC, rossi2014montecarlo} and studies of three-body scattering in vacuum \cite{braaten1999diluteBEC, kohler2002threebodyBEC, braaten2002diluteBEC, tan2008hypervolume}. Specifically,
zero-energy three-body collisions determine the $\text{ln}(n a^3)$ correction calculated by Wu \cite{wu1959energyBEC, sawada1959energyBEC, hugenholtz1959energyBEC} and the \textit{scattering hypervolume} $D$ \cite{tan2008hypervolume}.
Crucially, $D$ determines the effective three-body interaction in an analogous role to $a$ in the two-body case. It is predicted to act as a stabilizing force for quantum droplets in ultracold Bose gases \cite{bulgac2002droplets, bedaque2003droplets, blakie2016dropletsDipolar}, may be tuned experimentally \cite{petrov2014control3bodyInteractions}, and could be experimentally determined from the compressibility or sound modes of Bose gases \cite{zwerger2019phasetransition}.

The imaginary part of $D$ is proportional to the three-body recombination rate \cite{tan2017hypervolume,braaten2006universality} and has been studied extensively for various three-body systems, both experimentally and theoretically \cite{greene2017review}.
However, despite its fundamental relevance, the real part of $D$ remains sparsely explored. This is partly caused by the difficulty of removing singular contributions to the elastic three-body scattering amplitude required to obtain the real part of the scattering hypervolume \cite{amado1970elastic3bodyT, braaten1999diluteBEC, braaten2006universality, dincao2018review}.
In the strongly interacting regime, Efimov physics plays a dominant role leading to universal log-periodic behavior of $D$ \cite{efimov1979threebody, braaten2002diluteBEC, braaten2006universality, dincao2018review}. In the weakly interacting regime, $D$ has been studied considering the repulsive hard-sphere potential \cite{tan2008hypervolume} and a Gaussian interaction potential \cite{tan2017hypervolume}. However, the behavior of $D$ over a full range of interaction strengths has not been explored for any finite-range potential, which demonstrates the nontrivial character of this problem.

%Despite their fundamental relevance, low-energy elastic three-body scattering \ins{properties remain} sparsely explored. This is partly caused by the difficulty of removing singular contributions to the elastic three-body scattering amplitude required to obtain the hypervolume \cite{amado1970elastic3bodyT, braaten1999diluteBEC, braaten2006universality, dincao2018review}.
%The imaginary part of $D$ is proportional to the three-body recombination rate \cite{tan2017hypervolume,braaten2006universality} which has been studied extensively for various three-body systems, both experimentally and theoretically \cite{greene2017review}.
%Universal behavior has been predicted for the strongly interacting regime where Efimov physics plays a dominant role leading to a log-periodic behavior of $D$ \cite{efimov1979threebody, braaten2002diluteBEC, braaten2006universality, dincao2018review}. For short-range pairwise potentials, $D$ has been studied considering the repulsive hard-sphere potential \cite{tan2008hypervolume} and a Gaussian interaction potential \cite{tan2017hypervolume}. However, the behavior of $D$ over a full range of interaction strengths has not been explored for any finite-range potential, which demonstrates the highly nontrivial character of this problem.

In this Rapid Communication, we investigate the three-body scattering hypervolume $D$ for identical bosons interacting via a pairwise square-well potential, covering the range from weak to strong interactions, and analyze the corresponding universal and nonuniversal effects. We present the first numerical calculations of $D$ in the strongly interacting regime for a local finite-range potential, and study the corresponding Efimov universality. Besides the Efimov resonances, we identify additional three-body resonances close to two-body $d$- and $g$-wave resonances and study their character.

\textit{Elastic three-body scattering amplitude.}---A convenient way to calculate $D$ is to use the Faddeev equations for the three-particle transition operators $U_{\alpha \beta}$ in the form presented by Alt, Grassberger, and Sandhas (AGS) \cite{alt1967ags, schmid1974threebody},
\begin{equation} \label{eq:AGS_elastic}
\begin{cases}
U_{0 0}(z) = \sum_{\alpha = 1}^{3} T_{\alpha}(z) G_0(z) U_{\alpha 0}(z), \\
\begin{aligned}
U_{\alpha 0}(z) &= G_0^{-1}(z) + \sum_{\substack{\beta = 1 \\ \beta \neq \alpha}}^{3} T_{\beta}(z) G_0(z) U_{\beta 0}(z)
\\
& \text{ for } \alpha = 1, 2, 3,
\end{aligned}
\end{cases}
\end{equation}
to find the transition amplitude for three-body elastic scattering that is described by the operator $U_{0 0}(z)$. Here $z$ is the (complex) three-body energy. The index $\alpha$ ($\beta$) in $U_{\alpha \beta}(z)$ labels the four possible configurations for the outgoing (incoming) state of the three-body scattering wave function, i.e., $\alpha = 0$ denotes three free particles, whereas $\alpha = 1$, 2, and 3 stand for the three possible atom-dimer configurations.
$T_{\alpha}(z)$ represents the transition operator for scattering between particles $\beta$ and $\gamma$ ($\beta,\gamma = 1,2,3$, $\beta \neq \gamma \neq \alpha$) in the presence of particle $\alpha$ and is simply related to the two-body $T$ operator $T(z_{2b})$ \cite{mestrom2019squarewell}, where $z_{2b}$ is some complex value for the energy of the two-body system. The operator $G_0(z)$ is the free three-body Green's function $(z - H_0)^{-1}$ where $H_0$ is the three-body kinetic energy operator in the center-of-mass frame of the three-particle system.

The three-body configuration is parametrized by the Jacobi momenta $\mathbf{p}_{\alpha} = ( \mathbf{P}_{\beta} - \mathbf{P}_{\gamma})/2$ and $\mathbf{q}_{\alpha}~=~(2/3) \left[ \mathbf{P}_{\alpha} - \left(\mathbf{P}_{\beta} + \mathbf{P}_{\gamma}\right)/2\right]$ where $\mathbf{P}_{\alpha}$ represents the momentum of particle $\alpha$ in the laboratory frame.
There exist three possibilities to choose these Jacobi vectors. If we define $\mathbf{q} \equiv \mathbf{q}_1$ and $\mathbf{p} \equiv \mathbf{p}_1$, we have $\mathbf{q}_2 = \mathbf{p} - \mathbf{q}/2$, $\mathbf{p}_2 = -\mathbf{p}/2 - 3 \mathbf{q}/4$, $\mathbf{q}_3 = -\mathbf{p} - \mathbf{q}/2$, and $\mathbf{p}_3 = -\mathbf{p}/2 + 3 \mathbf{q}/4$.
This parametrization is suitable for relating the matrix element $\langle \mathbf{p}, \mathbf{q} | U_{0 0}(0) \lvert \mathbf{0}, \mathbf{0} \rangle$ to the scattering hypervolume $D$ where we normalize the plane-wave states according to $\langle \mathbf{p}' | \mathbf{p} \rangle = \delta(\mathbf{p}' - \mathbf{p})$.
From Tan's definition of the three-body scattering hypervolume $D$ \cite{tan2008hypervolume}, we deduce the following relation between $\langle \mathbf{p}, \mathbf{q} | U_{0 0}(0) \lvert \mathbf{0}, \mathbf{0} \rangle$ and $D$ (see Supplemental Material \cite{SupplMat}):
\begin{widetext}
\begin{eqnarray}\label{eq:U00_vs_D}
\begin{aligned}
\langle \mathbf{p}, \mathbf{q} | U_{0 0}(0) \lvert \mathbf{0}, \mathbf{0} \rangle = &\sum_{\alpha = 1}^{3} 
\Bigg\{
\delta(\mathbf{q}_{\alpha}) \langle \mathbf{p}_{\alpha} | T(0) | \mathbf{0}\rangle 
 - \frac{1}{2 \pi^4} \frac{a^2}{m \hbar^2} \frac{1}{q_{\alpha}^2}
+ \frac{1}{12 \pi^4}(4 \pi - 3 \sqrt{3}) \frac{a^3}{m \hbar^3} \frac{1}{q_{\alpha}}
+ \frac{1}{3 \pi^5}(4 \pi - 3 \sqrt{3}) \frac{a^4}{m \hbar^4} \, \text{ln}\bigg( \frac{q_{\alpha} |a|}{\hbar} \bigg)
\\
&
-\frac{p_{\alpha}^2 + \frac{3}{4} q_{\alpha}^2}{q_{\alpha}^2} 
\frac{a}{2 \pi^2 \hbar}  
\frac{\partial^2 \langle \mathbf{p} | T(0) | \mathbf{0}\rangle }{\partial p^2}\Bigg|_{p=0}
+ \frac{1}{3}\frac{1}{(2 \pi)^6}  \frac{D}{m \hbar^4} 
+ O\left(q_{\alpha} \, \text{ln}\bigg( \frac{q_{\alpha} |a|}{\hbar} \bigg), \frac{p_{\alpha}^2}{q_{\alpha}} \right)
\Bigg\},
\end{aligned}
\end{eqnarray}
\end{widetext}
which holds for any local symmetric two-body potential. Here the three-body energy $z = 0$ is approached from the upper half of the complex energy plane, which fixes the sign of the imaginary part of $D$.

We consider three identical bosons that interact via a pairwise square-well potential
\begin{equation}
V(r)=\begin{cases}-V_0,&\mbox{$0\leq r<R$},\\0,&\mbox{$r\geq R$},\end{cases}
\end{equation}
where $r$ denotes the relative distance between two particles and $R$ and $V_0$ represent the range and depth of the potential, respectively. To obtain $D$, we solve the AGS equations given in \refEquation{eq:AGS_elastic} for the matrix element $\langle \mathbf{p}, \mathbf{q} | U_{0 0}(0) | \mathbf{0}, \mathbf{0} \rangle$ after subtracting the terms in \refEquation{eq:U00_vs_D} that diverge as $p,q \to 0$ (see Ref.~\cite{SupplMat}). The dimension of this set of integral equations is reduced to one by expanding this amplitude in spherical harmonics and in two-body states that are determined by the Weinberg expansion and are thus related to two-body bound states or resonances \cite{weinberg1963expansion, mestrom2019squarewell}. The resulting integral equation is solved as a matrix equation by discretizing the momenta. 

Our method differs from another approach recently presented by Zhu and Tan \cite{tan2017hypervolume} who calculated the scattering hypervolume $D$ from the zero-energy three-body scattering wave function in position space for a variable two-body Gaussian potential. 
Their numerics were limited to the weakly interacting regime in contrast to our approach covering the complete regime ranging from strongly repulsive to attractive potentials and from weak ($|a|/R \lesssim 1$) to strong ($|a|/R \gg 1$) interactions. In the following, we show our results in these regimes obtained by tuning the potential depth $V_0$.

%We have calculated the three-body scattering hypervolume $D$ in the complete regime ranging from strongly repulsive potentials, i.e., $V_0 \to -\infty$, towards attractive potentials supporting multiple two-body bound states as shown in \refFigure{fig:SqW_Dh_MultipleRes}, and we have explored both the weakly- and strongly interacting regime, i.e., \ins{$|a|/R \lesssim 1$ and $|a|/R \gg 1$}, respectively, by tuning the potential depth $V_0$.

\begin{figure*}[hbtp]
    \centering
    \includegraphics[width=0.99\textwidth]{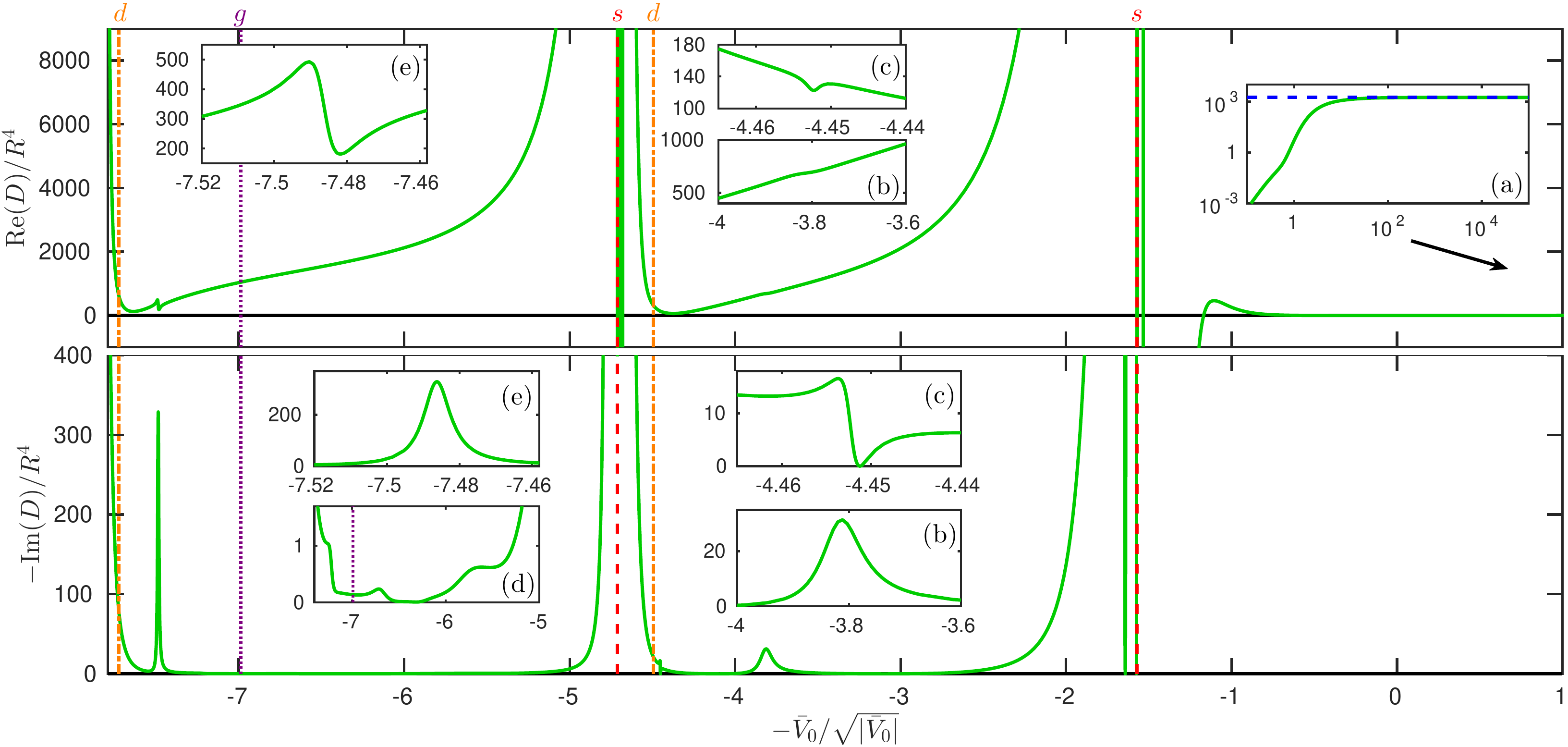}
    \caption{Three-body scattering hypervolume $D$ (green solid line) corresponding to the square-well potential as a function of the dimensionless interaction strength $\bar{V}_0 = m V_0 R^2/\hbar^2$. The vertical lines indicate the interaction strengths at which two-body states become bound: $s$-wave states ($l = 0$, red dashed lines) at $\bar{V}_0 = \left(\pi/2 \right)^2$ and $\left(3 \pi/2 \right)^2$, $d$-wave states ($l = 2$, orange dash-dotted lines) at $\bar{V}_0 \approx (4.49)^2$ and $(7.73)^2$, and $g$-wave state ($l = 4$, purple dotted line) at $\bar{V}_0 \approx (6.99)^2$. Inset (a) displays the behavior of $D$ for strongly repulsive potentials as indicated by the black arrow. The horizontal blue dashed line represents the hard-sphere limit calculated by Ref.~\cite{tan2008hypervolume}. The other insets (b)--(e) zoom in on the real and imaginary parts of $D$ near several resonances that arise from three-body quasibound states at the three-particle threshold.}
    \label{fig:SqW_Dh_MultipleRes}
\end{figure*}
% Figure made with the code: Plot_and_fit_Elastic_K3_List_v8_SqW_MultipleRes_Insets_h.m

\textit{Repulsive potentials.}---In the limit $V_0 \to -\infty$, the square-well potential approaches the hard-sphere interaction that was considered already a decade ago by Tan \cite{tan2008hypervolume}. Our results for $D$ in this limit are shown in \refFigure{fig:SqW_Dh_MultipleRes}(a) where we find good agreement within the numerical accuracy of our approach:
\begin{equation}
D/a^4 \underset{\bar{V}_0 \to -\infty}{=}  1761 \pm 1.
\end{equation}
%\begin{equation}
%D/a^4 = 1761 \pm 1 \quad \text{(hard-sphere limit)}.
%\end{equation}

When the potential barrier $-V_0$ is decreased, the scattering hypervolume decreases as well, and it eventually goes to zero in the limit $|V_0| \to 0$ as
\begin{align}
D &=  -96 \pi^6 m^2 \hbar^4 \langle \mathbf{0} | V | \mathbf{0} \rangle \frac{\partial^2 \langle \mathbf{p} | V | \mathbf{0} \rangle }{\partial p^2}\Bigg|_{p = 0} 
+ O(\bar{V}_0^3) \label{eq:D_small_V0_general}
\\
&= \frac{8}{15} \pi^2 \bar{V}_0^2 R^4 + O(\bar{V}_0^3), \label{eq:D_small_V0_SqW}
\end{align}
where $\bar{V}_0 = m V_0 R^2/\hbar^2$ denotes the dimensionless interaction strength. Equation~(\ref{eq:D_small_V0_general}) is a general relation for $D$ for local symmetric potentials $V$ in the zero-depth limit \cite{GeneralCaseZeroDepthLimit}, whereas \refEquation{eq:D_small_V0_SqW} applies specifically to the square-well potential. We have analytically derived \refEquation{eq:D_small_V0_general} from the AGS equations using the Born approximation (see Ref.~\cite{SupplMat}), and we have numerically confirmed it for the square-well potential. An expression equivalent to \refEquation{eq:D_small_V0_general} has been derived in position space by Ref.~\cite{tan2017hypervolume} that confirmed it for a Gaussian potential.

\textit{Attractive potentials.}---As the potential depth increases, two-body states start to become bound resulting in a nonzero value for the imaginary part of $D$. Figure~\ref{fig:SqW_Dh_MultipleRes} shows that this value is much smaller than the magnitude of the real part in most regimes. Close to the two-body $s$-wave potential resonances that are indicated by the vertical red dashed lines in \refFigure{fig:SqW_Dh_MultipleRes}, the pairwise interactions are strong ($|a|\gg R$). Here the scattering hypervolume $D$ scales as $a^4$ and its behavior becomes log-periodic due to the Efimov effect as we will see below. In between these two-body resonances where $|a| \lesssim R$, we identify several resonances related to three-body quasibound states that appear at the zero-energy threshold as indicated by Figs.~\ref{fig:SqW_Dh_MultipleRes}(b)--\ref{fig:SqW_Dh_MultipleRes}(e). In the following paragraphs, we first analyze the characteristics of these three-body resonances before presenting our results in the strongly interacting regime.
% We focus first on the behavior of $D$ in between the potential resonances where $|a| \lesssim R$. In this weakly interacting regime, we identify several resonances related to three-body quasibound states that appear at the zero-energy threshold as indicated by the insets in \refFigure{fig:SqW_Dh_MultipleRes}.}

%Besides the Efimov trimer states in the strongly interacting regime, additional three-body quasibound states might appear at the zero-energy threshold in the weakly interacting regime. 
%\ins{We will discuss this strongly interacting regime further below, but first we analyze the features in between the potential resonances where $|a| \lesssim R$.} \ins{We identify several three-body resonances related to such trimer states as indicated by the insets in \refFigure{fig:SqW_Dh_MultipleRes}}
%Additional three-body quasibound states might appear at the zero-energy threshold in the weakly interacting regime. \del{????????} We identify several three-body resonances related to such trimer states as indicated by the insets in \refFigure{fig:SqW_Dh_MultipleRes}.

%Besides the Efimov trimer states in the strongly interacting regime, additional three-body quasibound states might appear at the zero-energy threshold in the weakly interacting regime. \del{????????} We identify several three-body resonances related to such trimer states as indicated by the insets in \refFigure{fig:SqW_Dh_MultipleRes}. 

The presence of the trimer resonances at $\sqrt{\bar{V}_0} = 3.8$ and $4.45$ [Figs.~\ref{fig:SqW_Dh_MultipleRes}(b) and \ref{fig:SqW_Dh_MultipleRes}(c), respectively] depends critically on the inclusion of 
the almost bound two-body $d$-wave state (vertical orange dash-dotted line at $\sqrt{\bar{V}_0} \approx 4.49$) in our Weinberg expansion of the two-body $T$ operator. This suggests that these trimer states are associated with this $d$-wave dimer state in a similar way as the three-body state for van der Waals potentials studied by Ref.~\cite{dincao2012dwave}.

In \refFigure{fig:SqW_Dh_MultipleRes}(d) we highlight small features at $\sqrt{\bar{V}_0} = 5.7$, $6.7$, and $7.2$. These are close to the point at which the first $g$-wave dimer state gets bound (vertical purple dotted line). By analyzing the eigenvalues of the kernel of the integral equation (see Ref.~\cite{SupplMat}), we find that they are true trimer resonances. These resonances vanish when the first $g$-wave dimer state is removed from the Weinberg expansion. We do not see any effects of the resonances at $\sqrt{\bar{V}_0} = 5.7$, $6.7$, and $7.2$ on the real part of $D$ within our numerical accuracy. More generally, our results suggest that trimer resonances in the weakly interacting regime have a stronger effect on $\text{Im}(D)$ than on $\text{Re}(D)$.

The next trimer resonance occurs
at $\sqrt{\bar{V}_0} = 7.49$ [see \refFigure{fig:SqW_Dh_MultipleRes}(e)]. It vanishes when removing the second $d$-wave or the second $g$-wave dimer state from our Weinberg expansion. These two-body states get bound at $\sqrt{\bar{V}_0} \approx 7.73$ (vertical orange dash-dotted line) and $\sqrt{\bar{V}_0} \approx 10.42$, respectively. So both $d$-wave and $g$-wave effects play a significant role for this trimer resonance.

%Finally, we find three small resonances near the point at which the first $g$-wave dimer state gets bound (vertical purple dotted line).} \del{By analyzing the eigenvalues of the three-body kernel (see Ref.~\cite{SupplMat}), we find that they are true trimer resonances.} \ins{These resonances when the first $g$-wave dimer state is removed by excluding the corresponding expansion term in our calculation, which shows the strong $g$-wave dimer state. }

%The presence of these three resonances depends critically on the inclusion of one term in our Weinberg expansion of the two-body $T$ operator, namely the one corresponding to the almost bound two-body $d$-wave state (vertical orange dash-dotted lines). Therefore we conclude that these three trimer states are associated with $d$-wave dimer states in a similar way as the universal three-body state for van der Waals potentials studied by Ref.~\cite{dincao2012dwave}.

%\ins{Even though the trimer resonances at $\sqrt{\bar{V}_0} = 3.8$ and $4.45$ have the same origin,} 
%\refFigure{fig:SqW_Dh_MultipleRes} 
%\ins{Even though some trimer resonances seem to have a similar origin,} 
%\ins{Even though the trimer resonances are strongly related to a few dimer states,}

Even though these trimer resonances in the weakly interacting regime all originate from the nonzero partial-wave components of the two-body interaction potential, the behavior of $D$ is not the same for all resonances. This suggests that the behavior of the scattering hypervolume in the weakly interacting regime depends on some three-body background phase shift resulting from nonresonant pathways for three-body scattering \cite{dincao2018review}. 
%In particular, the line shape of the trimer resonance at $\sqrt{\bar{V}_0} = 4.45$ deviates from the other resonances. 
In particular, \refFigure{fig:SqW_Dh_MultipleRes}(c) shows a sharp minimum in $-\text{Im}(D)$ (or equivalently in the three-body recombination rate) near the trimer resonance peak at $\sqrt{\bar{V}_0} = 4.45$. Such a sharp feature was also encountered by Ref.~\cite{tan2017hypervolume} for a Gaussian potential supporting two $s$-wave dimer states. We suspect that both features arise from destructive interference effects \cite{dincao2018review}, since the minimum in \refFigure{fig:SqW_Dh_MultipleRes}(c) vanishes when we exclude the almost bound two-body $d$-wave state in our Weinberg expansion.

Our results presented in \refFigure{fig:SqW_Dh_MultipleRes} can be compared to the calculations of Ref.~\cite{tan2017hypervolume} for the scattering hypervolume corresponding to a Gaussian two-body potential. Even though both results are very similar for repulsive potentials, they are quite different for attractive interactions.
The main difference is the behavior of $D$ when approaching the $s$-wave dimer resonances (vertical red dashed lines), where Ref.~\cite{tan2017hypervolume} finds additional trimer resonances that are different from the Efimov resonances. 
%It is unclear where this difference comes from. 
%However, the validity of the method presented in Ref.~\cite{tan2017hypervolume} breaks down when $|a|/R \gg 1$, which makes the comparison in that regime questionable.
Secondly, we find that $D$ behaves smoothly across the $d$-wave dimer resonances (vertical orange dash-dotted lines) in contrast to the results of Ref.~\cite{tan2017hypervolume}. These differences show that the details of the considered two-body potential play a crucial role in the behavior of $D$ across a $d$-wave dimer resonance and on the presence of trimer states in the weakly interacting regime. 
%This difference suggest that the behavior of $D$ across a $d$-wave dimer resonance depends on the details of the considered two-body potential. 
%For example, in case of van der Waals potentials it has been shown that the three-body recombination rate (or equivalently $-\text{Im}(D)$) is strongly enhanced when the $d$-wave dimer state becomes bound due to the opening up of an additional decay channel for three free particles \cite{dincao2012dwave}.

We now discuss our results in the strongly interacting regime ($|a|/R \gg 1$). Here, the behavior of $D$ is predicted to follow a general form determined in Refs.~\cite{efimov1979threebody,braaten2002diluteBEC,esry1999recombination,braaten2000K3, braaten2001K3deep} and generalized in Refs.~\cite{braaten2004atomdimer, braaten2006universality, dincao2018review} by including the inelasticity parameter $\eta$ that describes the tendency to decay to deeply bound dimer states. 
These limiting forms for $D$ contain a number of universal constants obtained in Refs.~\cite{petrov2005unpublished, macek2006analytical, gogolin2008analytical, braaten2004atomdimer, braaten2002diluteBEC} which we refine in this work (see Ref.~\cite{SupplMat}). In addition to $\eta$, they also depend on the nonuniversal parameters $a_-$ and $a_+$ that locate the three-body recombination maxima and minima, respectively, and are completely determined by the interaction between the three particles \cite{braaten2006universality}.

%\ins{Universal relations for the three-body recombination rate which is proportional to $\text{Im}(D)$ have been determined in Refs.~\cite{braaten2000K3, braaten2001K3deep, braaten2004atomdimer, braaten2006universality}. }
%We now discuss our results in the strongly interacting regime ($|a|/R \gg 1$). Here, the behavior of $D$ is predicted to follow a general form derived in Refs. ??? and generalized in Ref. ?? by including the inelasticity parameter $\eta$ that describes the tendency to decay to deeply bound dimer states. These limiting forms contain a number of universal constants obtained in Refs. ??? which we refine in this work (see supp???). In addition to $\eta$, they also depend on the nonuniversal parameters $a_-$ and $a_+$ that locate the three-body recombination maxima and minima, respectively, and are completely determined by the interaction between the three particles \cite{braaten2006universality}.

The universal expressions for the real part of the scattering hypervolume $D$ are given by
\begin{eqnarray} \label{eq:Dhyp-_universal_with_eta}
\text{Re}(D/a^4) &\approx C \Bigg(c_- 
+ \frac{ \frac{1}{2} b_- \sin \Big(2 s_0 \, \text{ln}(a/a_-) \Big)}{\sin^2 \Big( s_0 \, \text{ln}(a/a_-) \Big) + \sinh^2(\eta)}
\Bigg)
\end{eqnarray}
for $a<0$ and 
\begin{equation} \label{eq:Dhyp+_universal_with_eta}
\begin{aligned}
\text{Re}(D/a^4) &\approx C \Bigg(c_+ + \frac{1}{2} b_+ (1 - e^{-2 \eta}) \\
&+ b_+ e^{-2 \eta} \sin^2\Big( s_0 \ \text{ln}(a/a_+) - \pi/4 \Big)\Bigg)
\end{aligned}
\end{equation}
for $a>0$. 
The imaginary part of $D$ is given by the universal formulas
\begin{equation}\label{eq:ImDhyp-_universal_with_eta}
\text{Im}(D/a^4) \approx -\frac{1}{2} C_{-} \frac{\sinh(2 \eta)}{\sin^2\left(s_0 \ln(a/a_{-})\right) + \sinh^2(\eta)} 
\end{equation}
for $a<0$ and
\begin{equation} \label{eq:ImDhyp+_universal_with_eta}
\begin{aligned}
\text{Im}&(D/a^4) \approx -\frac{1}{2} C_{+} \Bigg( \frac{1}{4}  \left(1-e^{-4 \eta} \right) 
\\
&
+e^{-2 \eta} \bigg(\sin^2\left(s_0 \ln(a/a_{+})\right) + \sinh^2(\eta) \bigg)
\Bigg)
\end{aligned}
\end{equation} 
for $a>0$.
Here $s_0 \approx 1.00624$ is the constant that sets the periodicity in Efimov physics for identical bosons \cite{efimov1970energy, efimov1971weakly} and we have defined the constant $C \equiv 64 \pi (4 \pi - 3 \sqrt{3})$. The coefficients $b_{\pm}$, $c_{\pm}$, and $C_{\pm}$ are universal in the sense that they do not depend on the short-range form of the potentials \cite{dincao2018review}. These constants were determined previously to be $C_- \approx 4590$ \cite{braaten2004atomdimer}, $C_+ \approx 67.1177$ \cite{petrov2005unpublished, macek2006analytical, gogolin2008analytical}, $b_- = 3.16$, $c_- = 1.14$, $b_+ = 0.021$, and $c_+ = 1.13$ \cite{braaten2002diluteBEC} (see Ref.~\cite{SupplMat} for the connection between $D$ and the quantity calculated in Ref.~\cite{braaten2002diluteBEC}).

\begin{figure*}[hbtp]
	\begin{subfigure}
	\centering
	\includegraphics[width=3.4in]{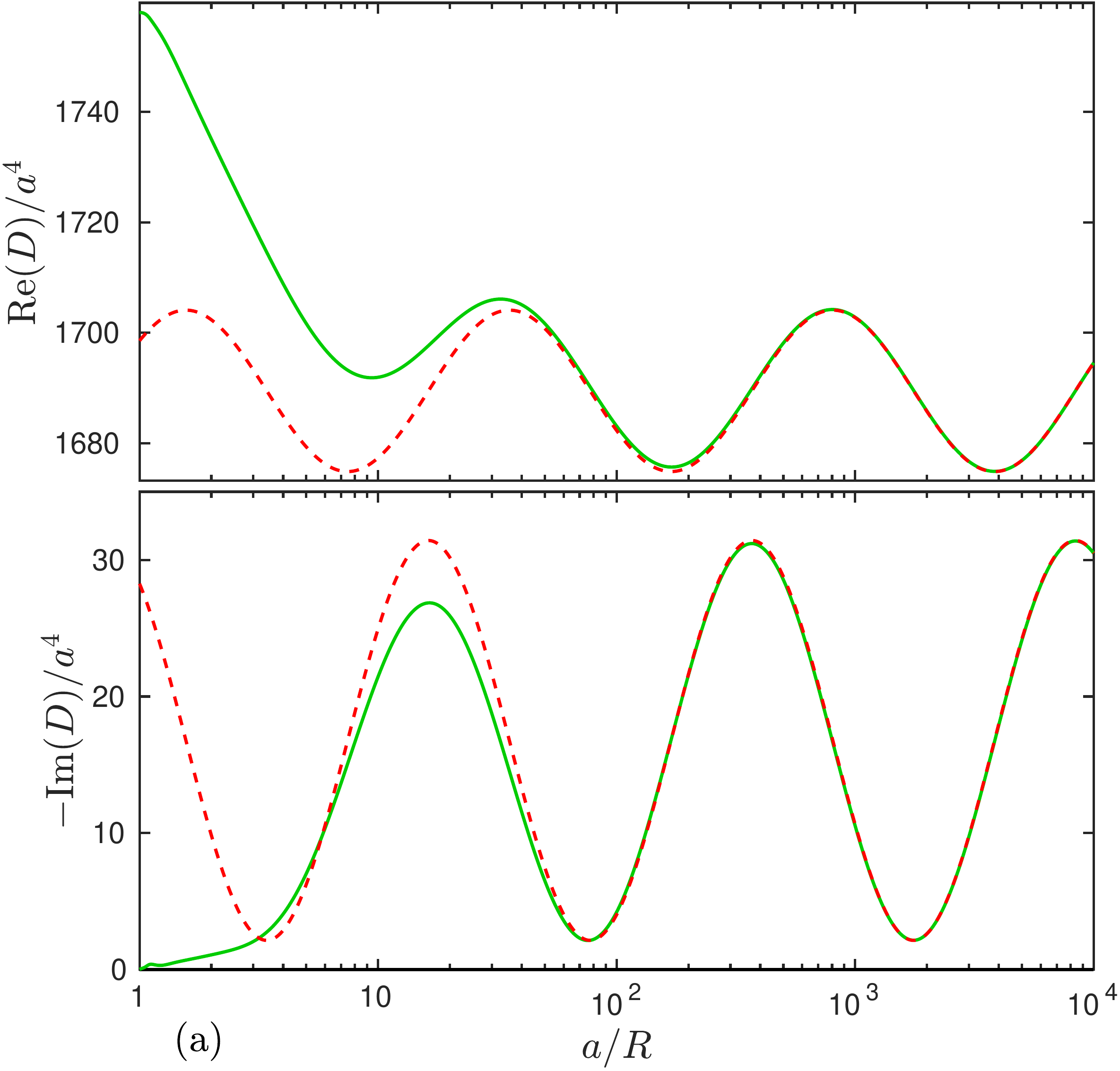}
	\end{subfigure}
\quad
	\begin{subfigure}
	\centering
	\includegraphics[width=3.4in]{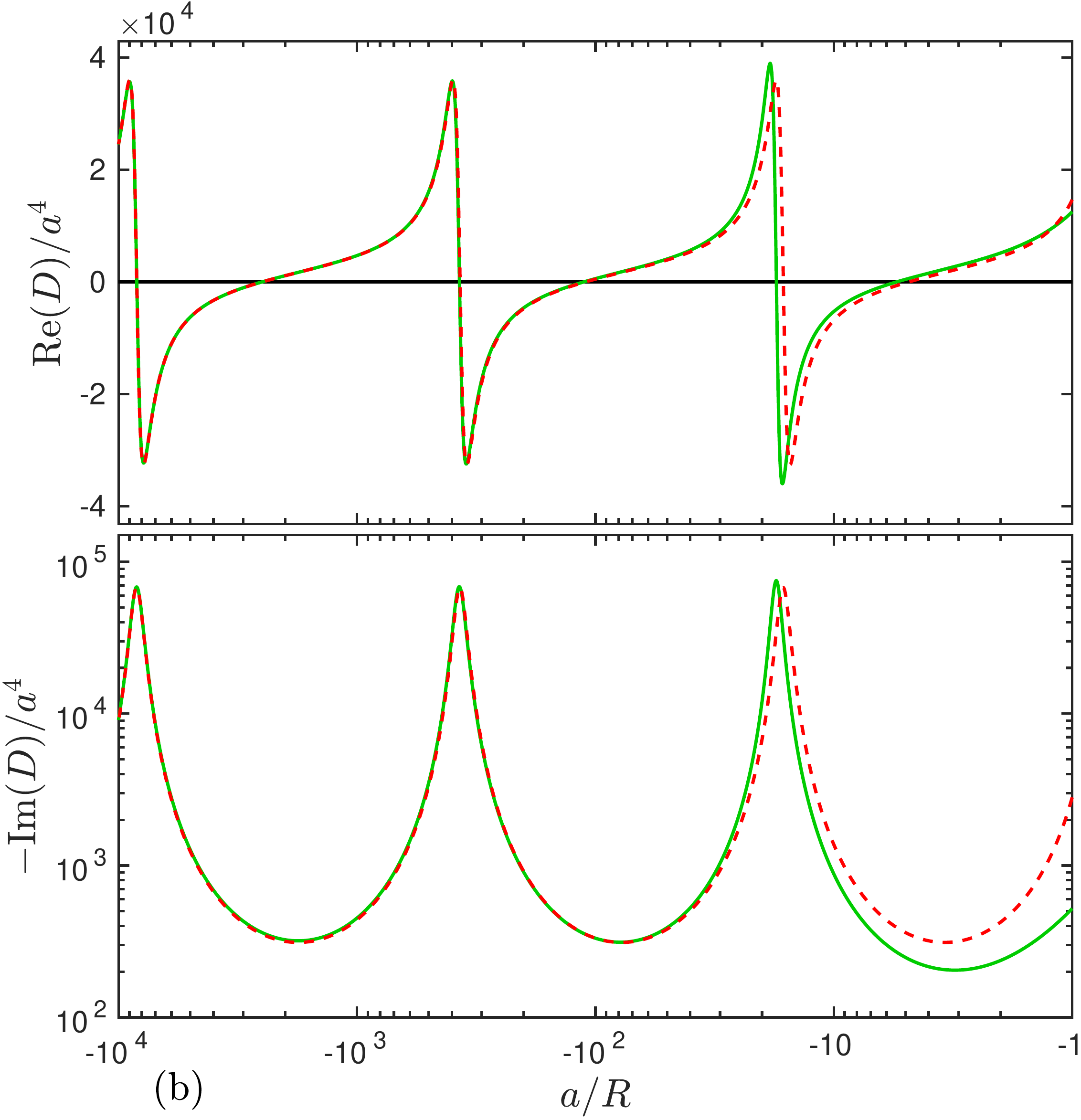}
	\end{subfigure}
    \caption{Three-body scattering hypervolume $D$ (green solid line) near the second potential resonance of the square-well potential for (a) $a>0$ and (b) $a<0$. The dashed curves give the analytic zero-range results given by Eqs.~(\ref{eq:Dhyp-_universal_with_eta}), (\ref{eq:Dhyp+_universal_with_eta}), (\ref{eq:ImDhyp-_universal_with_eta}) and (\ref{eq:ImDhyp+_universal_with_eta}) where we set $a_+/R = 1759$, $b_+ = 0.0226$, $c_+ = 1.1288$, $C_+ = 67.118$, $a_-/R = -8396$, $b_- = 3.153$, $c_- = 1.140$, $C_- = 4590$ and $\eta = 0.068$.}
    \label{fig:SqW_Res2_Dhyp_a0_neg_and_pos}
\end{figure*}
% Figure made with code Plot_and_fit_Elastic_K3_List_v9_SqW_Res2_a0neg_b_v3_No_fit_Im.m AND Plot_and_fit_Elastic_K3_List_v9_SqW_Res2_a0neg_b_v3_No_fit_Im.m

We have redetermined the universal coefficients as $C_+ = 67.118(5)$, $b_+ = 0.0226(5)$, $c_+ = 1.1288(5)$, $b_- = 3.153(5)$, and $c_- = 1.140(2)$. This was done by analyzing the three-body scattering hypervolume for a contact interaction with a cutoff in momentum space (see Ref.~\cite{SupplMat} for details). We find good agreement with the previously determined values except for $b_+$ that deviates approximately $7\%$ from Ref.~\cite{braaten2002diluteBEC}. However, this leads only to a deviation of $0.1\%$ in the overall value of $D$ (see Ref.~\cite{SupplMat}).

%Our value for $C_+$ is in excellent agreement with the analytical prediction \cite{petrov2005unpublished, macek2006analytical, gogolin2008analytical}.

The universal relations (\ref{eq:Dhyp-_universal_with_eta}) and (\ref{eq:Dhyp+_universal_with_eta}) have not been previously tested numerically for any local nonzero-range two-body potential. Near the second two-body $s$-wave potential resonance of the square-well potential, i.e., $\sqrt{\bar{V}_0}$ is close to $3 \pi/2$, we compare our results against the universal relations in \refFigure{fig:SqW_Res2_Dhyp_a0_neg_and_pos}. Using our results for the universal constants, we numerically confirm Eqs.~(\ref{eq:Dhyp-_universal_with_eta}), (\ref{eq:Dhyp+_universal_with_eta}), (\ref{eq:ImDhyp-_universal_with_eta}), and (\ref{eq:ImDhyp+_universal_with_eta}). For this specific two-body resonance, we find that $a_{+}/R = 1759(5)$, $a_{-}/R = -8396(1)$, and $\eta = 0.068(1)$. Similar results for the first potential resonance can be found in the Supplemental Material \cite{SupplMat}.

%We compare our results in \refFigure{fig:SqW_Res2_Dhyp_a0_neg_and_pos} covering the universal regime near the second potential resonance of the square-well potential, i.e., $\sqrt{\bar{V}_0}$ is close to $3 \pi/2$.  
%The dashed curves display the analytic zero-range results given by Eqs.~(\ref{eq:Dhyp-_universal_with_eta}), (\ref{eq:Dhyp+_universal_with_eta}), (\ref{eq:ImDhyp-_universal_with_eta}) and (\ref{eq:ImDhyp+_universal_with_eta}). 
%The values for the nonuniversal parameters $a_{\pm}$ and $\eta$ are given in the Supplemental Material \cite{SupplMat} and are consistent with the results stated in Ref.~\cite{mestrom2019squarewell}. Similar results for the first potential resonance can be found in the Supplemental Material \cite{SupplMat}. 

%Our results for the coefficients $b_{\pm}$ and $c_{\pm}$ can be compared to the values presented by Braaten et al. \cite{braaten2002diluteBEC} who performed three-body calculations for a zero-range interaction. According to Ref.~\cite{braaten2002diluteBEC} the universal coefficients in Eqs. (\ref{eq:Dhyp-_universal_with_eta}) and (\ref{eq:Dhyp+_universal_with_eta}) are $b_- = 3.16$, $c_- = 1.14$, $b_+ = 0.021$ and $c_+ = 1.13$ (see Ref.~\cite{SupplMat}). Our values for $c_{\pm}$ and $b_{-}$ agree with the results of Ref.~\cite{braaten2002diluteBEC}. Even though our value of $b_+$ deviates approximately $7\%$ from the value of Ref.~\cite{braaten2002diluteBEC}, it is  only a deviation of $0.1\%$ from the overall result for $D$ (\del{see Fig.~S5 of Ref.~\cite{SupplMat}}).

\textit{Conclusion.}---By solving the AGS equations for the three-body elastic scattering amplitude, we have studied the behavior of the three-body scattering hypervolume $D$ which
is a fundamental quantity of ultracold three-body collisions and is needed for studying ultracold Bose gases beyond the LHY correction.
We have presented the first numerical calculations of $D$ for identical bosons with a variable nonzero-range potential in the strongly interacting regime. Our results agree with the universal predictions of Refs.~\cite{efimov1979threebody,braaten2002diluteBEC,esry1999recombination,braaten2000K3, braaten2001K3deep, braaten2004atomdimer, braaten2006universality, dincao2018review} and show how finite-range effects start to play a role as the absolute value of the scattering length is decreased. For repulsive interactions, we have confirmed the hard-sphere limit from Ref.~\cite{tan2008hypervolume} and the weak-interaction limit from Ref~\cite{tan2017hypervolume}.
We have also explored the weakly interacting regime for attractive potentials supporting up to two $s$-wave dimer states and identified several three-body resonances related to trimer states that depend strongly on $d$-wave and/or $g$-wave effects.
%at least one dimer state. We have identified several three-body resonances related to trimer states that depend strongly on $d$-wave and/or $g$-wave effects.} 

The approach outlined in this Rapid Communication is very general and can be applied to other types of two-body potentials as well, such as van der Waals potentials. It could also be applied to mixtures, for which low-energy elastic three-body scattering properties are completely unexplored. 
Additionally, one could extend our approach to study three-body scattering embedded in a many-body environment \cite{colussi2018dynamical} and determine how three-body correlations affect both stationary and dynamical observables of ultracold Bose gases for any short-range two-body potential. 
In particular, one could make quantitative predictions for the ground-state energy density of a BEC and investigate
stabilizing effects from the three-body scattering hypervolume for small negative scattering lengths including the formation of quantum droplets \cite{bulgac2002droplets, bedaque2003droplets, blakie2016dropletsDipolar}.

\textit{Acknowledgments.}---We thank Jos\'e P. D'Incao, Chris H. Greene, Silvia Musolino, Denise Braun, Gijs Groeneveld, and Jinglun Li for useful discussions. This research is financially supported by the Netherlands Organisation for Scientific Research (NWO) under Grant No. 680-47-623, and by the Foundation for Fundamental Research on Matter (FOM).

\bibliographystyle{apsrev}
\bibliography{Bibliography}

\begin{thebibliography}{71}
\bibitem[{Gen()}]{GeneralCaseZeroDepthLimit}
\bibinfo{note}{For this general case, $\bar{V}_0$ is defined to be a prefactor
  of the potential as $V(r) = \bar{V}_0 f(r)$ where $f(r)$ is independent of
  $\bar{V}_0$.}
\end{thebibliography}

%%%%%%%%%%%%%%%% SUPPLEMENTARY

\pagebreak
\clearpage % Needed because otherwise I did not get a new page

\onecolumngrid
\begin{center}
  \textbf{\large Supplemental Materials: ``Scattering hypervolume for ultracold bosons from weak to strong interactions''}\\[.2cm]
  P. M. A. Mestrom,$^{1}$ V. E. Colussi,$^{1}$ T. Secker,$^{1}$ and S. J. J. M. F. Kokkelmans$^1$\\[.1cm]
  {\itshape ${}^1$Eindhoven University of Technology, P.~O.~Box 513, 5600 MB Eindhoven, The Netherlands\\}
%(Dated: \today)\\[1cm]
\end{center}
\twocolumngrid

\setcounter{equation}{0}
\setcounter{figure}{0}
\setcounter{page}{1}
\renewcommand{\theequation}{S\arabic{equation}}
\renewcommand{\thefigure}{S\arabic{figure}}
\renewcommand{\thetable}{S\arabic{table}}  
\renewcommand{\bibnumfmt}[1]{[S#1]}

\section{The three-body scattering hypervolume}

Here we relate the three-body scattering hypervolume $D$ to the transition amplitude $\langle \mathbf{p}, \mathbf{q} | U_{0 0}(0) | \mathbf{0}, \mathbf{0} \rangle$ for identical bosons interacting via pairwise local symmetric potentials.
Tan \cite{tan2008hypervolume} defined the three-body scattering hypervolume $D$ via the zero-energy three-body scattering wave function $\lvert \Psi_{\text{3b}}(0) \rangle = \lvert \mathbf{0}, \mathbf{0} \rangle + G_0(0) U_{00}(0) \lvert \mathbf{0}, \mathbf{0} \rangle$, whose momentum-space representation is given by 
\begin{equation}\label{eq:Psi_3b_vs_U00}
\begin{aligned}
\langle \mathbf{p}, \mathbf{q} | \Psi_{\text{3b}} (0) \rangle 
&= \langle \mathbf{p} |  \mathbf{0} \rangle 
\langle \mathbf{q} |  \mathbf{0} \rangle
\\
&
- \frac{1}{\frac{p^2}{m} + \frac{3 q^2}{4 m}}
\langle \mathbf{p}, \mathbf{q} | U_{0 0}(0) \lvert \mathbf{0}, \mathbf{0} \rangle.
\end{aligned}
\end{equation}
In order to determine the behavior of the matrix element $\langle \mathbf{p}, \mathbf{q} | U_{0 0}(0) | \mathbf{0}, \mathbf{0} \rangle$, we analyze the AGS equations for identical bosons. As presented in Ref.~\cite{mestrom2019squarewell}, we define the operator $\breve{U}_{\alpha 0}(z) \equiv T_{\alpha}(z) G_0(z) U_{\alpha 0}(z) (1+P)$ where $P$ is the sum of the cyclic and anticyclic permutation operators. It satisfies the inhomogeneous equation
\begin{equation}\label{eq:Ubreve}
\breve{U}_{\alpha 0}(z) = T_{\alpha}(z)(1+P) + T_{\alpha}(z) G_0(z) P \breve{U}_{\alpha 0}(z).
\end{equation}
From \del{Eq.~(2) of the main text}, we derive
\begin{equation}\label{eq:U00_vs_Ubreve}
U_{00}(z) (1+P) = \sum_{\alpha = 1}^{3} \breve{U}_{\alpha 0}(z), 
\end{equation}
which gives
\begin{equation}\label{eq:U00_vs_Ubreve_matrix}
\langle \mathbf{p}, \mathbf{q} | U_{00}(z) | \mathbf{0}, \mathbf{0} \rangle = \frac{1}{3} \sum_{\alpha = 1}^{3} {}_{\alpha}\langle \mathbf{p}_{\alpha}, \mathbf{q}_{\alpha}| \breve{U}_{\alpha 0}(z) | \mathbf{0}, \mathbf{0} \rangle.
\end{equation}
The index $\alpha$ in $\lvert \mathbf{p}_{\alpha}, \mathbf{q}_{\alpha} \rangle_{\alpha}$ indicates that $\mathbf{p}_{\alpha}$ represents the relative momentum between particles $\beta$ and $\gamma$, whereas $\mathbf{q}_{\alpha}$ represents the relative momentum between particle $\alpha$ and the center of mass of the two-particle system $(\beta \gamma)$.

The singular behavior of ${}_{\alpha}\langle \mathbf{p}_{\alpha}, \mathbf{q}_{\alpha}| \breve{U}_{\alpha 0}(0) | \mathbf{0}, \mathbf{0} \rangle$ at $q_{\alpha}~=~0$ can be determined by writing the operator $\breve{U}_{\alpha 0}$ as a series and analyzing each term.
We write \refEquation{eq:Ubreve}  as
\begin{equation}\label{eq:Ubreve_series}
\begin{aligned}
\breve{U}_{\alpha 0} &= T_{\alpha} (1 + P) + T_{\alpha} G_0 P T_{\alpha} (1 + P)
\\
&
+ T_{\alpha} G_0 P T_{\alpha} G_0 P T_{\alpha} (1 + P)
+ ...
\end{aligned}
\end{equation}
where we removed the arguments of the operators for notational convenience.
The first term on the right-hand-side represents two-particle scattering in which the third particle only spectates,
\begin{equation}
\begin{aligned}
{}_{\alpha}\langle \mathbf{p}_{\alpha}, \mathbf{q}_{\alpha}| &T_{\alpha}(0) (1 + P) | \mathbf{0}, \mathbf{0} \rangle
\\
&= 3 \langle  \mathbf{q}_{\alpha} | \mathbf{0} \rangle
\langle \mathbf{p}_{\alpha} | T(0) | \mathbf{0} \rangle.
\end{aligned}
\end{equation}
The second term on the right-hand-side of \refEquation{eq:Ubreve_series} behaves as
\begin{equation}\label{eq:TG0PT(1+P)_matrix}
\begin{aligned}
{}_{\alpha} &\langle \mathbf{p}_{\alpha}, \mathbf{q}_{\alpha} | T_{\alpha}(0) G_0(0) P T_{\alpha}(0) (1 + P) | \mathbf{0}, \mathbf{0} \rangle
\\
&
\\
&= -\frac{3}{2 \pi^4} \frac{a^2}{m \hbar^2} \frac{1}{q_{\alpha}^2}
- 
\frac{3 \sqrt{3}}{4 \pi^4} \frac{a^3}{m \hbar^3} \frac{1}{q_{\alpha}}
\\
&
-3 m \langle \mathbf{0}| T(0) | \mathbf{0} \rangle 
\Bigg[
\Bigg(\frac{5}{4} + \frac{p_{\alpha}^2}{q_{\alpha}^2} \Bigg) \frac{\partial^2 \langle \mathbf{p}| T(0) | \mathbf{0} \rangle}{\partial p^2}\Bigg|_{p=0} 
\\
&+ \frac{\partial^2 \langle \mathbf{0}| T(-\frac{3}{4 m} q^2) | \mathbf{0} \rangle}{\partial q^2}\Bigg|_{q=0}
+ O(q_{\alpha}, p_{\alpha}^2/q_{\alpha})
\Bigg].
\end{aligned}
\end{equation}
The small-momentum behavior of the third term is given by
\begin{equation}\label{eq:TG0PTG0PT(1+P)_matrix}
\begin{aligned}
{}_{\alpha} &\langle \mathbf{p}_{\alpha}, \mathbf{q}_{\alpha} | \big(T_{\alpha}(0) G_0(0) P\big)^2 T_{\alpha}(0) (1 + P) | \mathbf{0}, \mathbf{0} \rangle
\\
&
\\
&= \frac{1}{\pi^3} \frac{a^3}{m \hbar^3} \frac{1}{q_{\alpha}} 
- \frac{3 \sqrt{3}}{\pi^5} \frac{a^4}{m \hbar^4} \text{ln}\Bigg(\frac{q_{\alpha} |a|}{\hbar}\Bigg)
\\
&
+ O\left(q_{\alpha}^0, \frac{p_{\alpha}^2}{q_{\alpha}}\right).
\end{aligned}
\end{equation}
It is a matter of choice to express the momentum $q_{\alpha}$ in units of $\hbar/|a|$ inside the logarithm. We choose these units to be consistent with the notation in Ref.~\cite{tan2008hypervolume}. The next term in the expansion of ${}_{\alpha}\langle \mathbf{p}_{\alpha}, \mathbf{q}_{\alpha}| \breve{U}_{\alpha 0}(0) | \mathbf{0}, \mathbf{0} \rangle$ also diverges logarithmically in the limit $q_{\alpha} \to 0$:
\begin{equation}
\begin{aligned}
{}_{\alpha} &\langle \mathbf{p}_{\alpha}, \mathbf{q}_{\alpha} | \big(T_{\alpha}(0) G_0(0) P\big)^3 T_{\alpha}(0) (1 + P) | \mathbf{0}, \mathbf{0} \rangle
\\
&
\\
&= \frac{4}{\pi^4} \frac{a^4}{m \hbar^4} \text{ln}\Bigg(\frac{q_{\alpha} |a|}{\hbar}\Bigg)
+ O\left(q_{\alpha}^0, p_{\alpha}^2\text{ln}(q_{\alpha}|a|/\hbar)\right).
\end{aligned}
\end{equation}
All other contributions to ${}_{\alpha}\langle \mathbf{p}_{\alpha}, \mathbf{q}_{\alpha}| \breve{U}_{\alpha 0}(0) | \mathbf{0}, \mathbf{0} \rangle$ are nonsingular in $q_{\alpha} = 0$. From the above analysis we find that $\langle \mathbf{p}, \mathbf{q} | \Psi_{\text{3b}} (0) \rangle$ defined by \refEquation{eq:Psi_3b_vs_U00} can be written as 
\begin{widetext}
\begin{equation}\label{eq:Psi3b_D}
\begin{aligned}
\langle \mathbf{p}, \mathbf{q} | \Psi_{\text{3b}} (0) \rangle 
&= \delta(\mathbf{p}) \delta(\mathbf{q})
- \frac{1}{\frac{p^2}{m} + \frac{3 q^2}{4 m}}
\sum_{\alpha = 1}^3 
\Bigg\{
\delta(\mathbf{q}_{\alpha}) \langle \mathbf{p}_{\alpha} | T(0) | \mathbf{0}\rangle 
 - \frac{1}{2 \pi^4} \frac{a^2}{m \hbar^2} \frac{1}{q_{\alpha}^2}
+ \frac{1}{12 \pi^4}(4 \pi - 3 \sqrt{3}) \frac{a^3}{m \hbar^3} \frac{1}{q_{\alpha}}
\\
&
+ \frac{1}{3 \pi^5}(4 \pi - 3 \sqrt{3}) \frac{a^4}{m \hbar^4} \, \text{ln}\bigg( \frac{q_{\alpha} |a|}{\hbar} \bigg)
-\frac{p_{\alpha}^2 + \frac{3}{4} q_{\alpha}^2}{q_{\alpha}^2} 
 \frac{a}{2 \pi^2 \hbar}  
\frac{\partial^2 \langle \mathbf{p} | T(0) | \mathbf{0}\rangle }{\partial p^2}\Bigg|_{p=0}
+ \frac{1}{3}\frac{1}{(2 \pi)^6}  \frac{D}{m \hbar^4} 
\\
&
+ O\left(q_{\alpha} \, \text{ln}\bigg( \frac{q_{\alpha} |a|}{\hbar} \bigg), \frac{p_{\alpha}^2}{q_{\alpha}} \right)
\Bigg\},
\end{aligned}
\end{equation}
\end{widetext}
where we used exactly the same definition for $D$ as the one presented by Tan \cite{tan2008hypervolume}. The relation between $D$ and  $\langle \mathbf{p}, \mathbf{q} | U_{00}(z) | \mathbf{0}, \mathbf{0} \rangle$ is given in \del{Eq.~(3) of the main text.} 

In order to see that \refEquation{eq:Psi3b_D} is the same as Eq.~(3) of Ref.~\cite{tan2008hypervolume}, we first relate the two-body transition amplitude $\langle \mathbf{p} | T(0) \lvert \mathbf{0} \rangle$ to the quantity $u_0$ as defined by Eq.~(2) of Ref.~\cite{tan2008hypervolume}. For this purpose, we note that the momentum-space representation of the zero-energy two-body scattering wave function $\lvert \Psi_{\text{2b}} (0) \rangle$ is given by
\begin{equation}
\begin{aligned}
\langle \mathbf{p} | \Psi_{\text{2b}} (0) \rangle 
&= \langle \mathbf{p} |  \mathbf{0} \rangle + \langle \mathbf{p} | G_0(0) T(0) \lvert \mathbf{0} \rangle
\\
&= \langle \mathbf{p} |  \mathbf{0} \rangle - \frac{m}{p^2}
\langle \mathbf{p} | T(0) \lvert \mathbf{0} \rangle
\\
&= \delta( \mathbf{p} ) - \frac{a_0}{2 \pi^2 \hbar} \frac{1}{p^2} 
\\
&
- \frac{1}{2} m \frac{\partial^2 \langle \mathbf{p} | T(0) | \mathbf{0}\rangle }{\partial p^2}\Bigg|_{p=0} 
+ O (p^2).
\end{aligned}
\end{equation}
Comparing this result to Eq.~(2) of Tan \cite{tan2008hypervolume}, we find that $u_0$ can be written as
\begin{equation} \label{eq:Def_u0}
u_0 = - 4 \pi^3 m \hbar^3 \frac{\partial^2 \langle \mathbf{p} | T(0) | \mathbf{0}\rangle }{\partial p^2}\Bigg|_{p=0} .
\end{equation}
Using this relation for $u_0$, \refEquation{eq:Psi3b_D} can be written in exactly the same form as Eq.~(3) of Ref.~\cite{tan2008hypervolume}.

\section{Scattering hypervolume for weak interaction strengths}

In this section we analyze the behavior of the three-body scattering hypervolume corresponding to three identical bosons interacting via a symmetric pairwise potential $V(r) = v_0 f(r)$ in the limit $v_0 \to 0$. Here $f(r)$ is some function independent of $v_0$ that goes to zero sufficiently fast for increasing interparticle separation such that regular scattering theory is valid \cite{taylor1972scattering}.

In the limit $v_0 \to 0$, we can approximate the two-body transition operator by the Born approximation $T(z) = V + O(v_0^2)$. To determine the scattering hypervolume $D$ to the lowest order in $v_0$, we consider the operator $T_{\alpha}(0) G_0(0) P T_{\alpha}(0) (1 + P)$. Applying the Born approximation to \refEquation{eq:TG0PT(1+P)_matrix}, we find
\begin{equation} 
\begin{aligned}
{}_{\alpha} &\langle \mathbf{0}, \mathbf{q}_{\alpha} | T_{\alpha}(0) G_0(0) P T_{\alpha}(0) (1 + P) | \mathbf{0}, \mathbf{0} \rangle
\\
&= -\frac{3}{2 \pi^4} \frac{a^2}{m \hbar^2} \frac{1}{q_{\alpha}^2}
- 
\frac{3 \sqrt{3}}{4 \pi^4} \frac{a^3}{m \hbar^3} \frac{1}{q_{\alpha}}
\\
&
-\frac{15 m}{4} \langle \mathbf{0}| V | \mathbf{0} \rangle 
 \frac{\partial^2 \langle \mathbf{q}| V | \mathbf{0} \rangle}{\partial q^2}\Bigg|_{q=0} 
+ O(q_{\alpha},v_0^3).
\end{aligned}
\end{equation}
Comparing this result with \del{Eq.~(3) of the main text} and using Eqs.~(\ref{eq:U00_vs_Ubreve_matrix}) and (\ref{eq:Ubreve_series}), we find that
\begin{equation}
\begin{aligned}
D =  -96 \pi^6 m^2 \hbar^4 \langle \mathbf{0} | V | \mathbf{0} \rangle \frac{\partial^2 \langle \mathbf{p} | V | \mathbf{0} \rangle }{\partial p^2}\Bigg|_{p = 0} 
+ O(v_0^3)
\end{aligned}
\end{equation}
as stated in the main text.

\section{Comparison with Braaten et al. \cite{braaten2002diluteBEC}}

\begin{figure}[hbtp]
    \centering
    \includegraphics[width=3.4in]{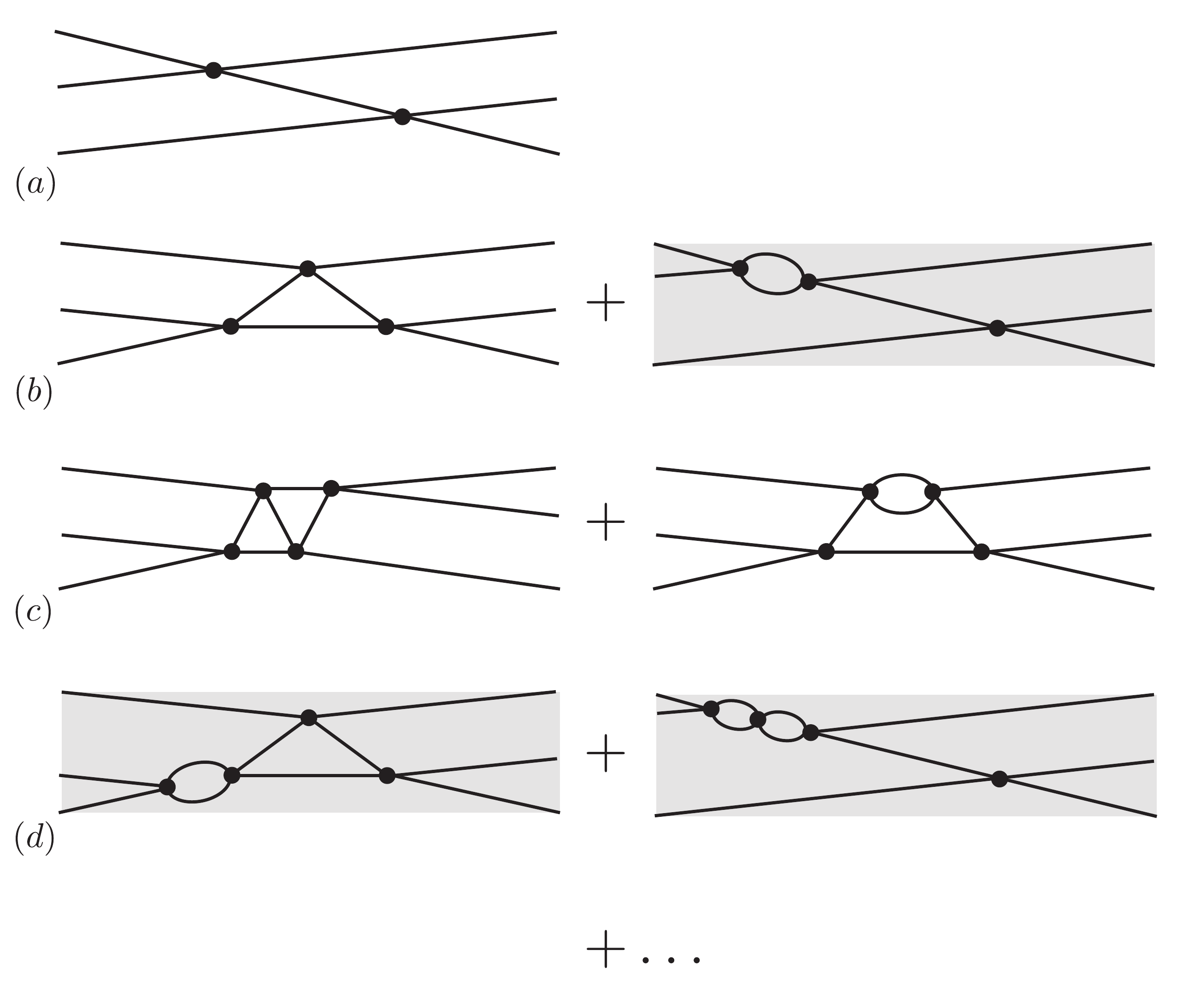}
    \caption{Feynman diagrams contributing to the $q_{\alpha}^{-2}$ behavior (a), the $q_{\alpha}^{-1}$ behavior (b), the $\text{ln}(q_{\alpha} |a|/\hbar)$ behavior (c) and the nondiverging part (d) of the three-body elastic scattering amplitude $\langle \mathbf{p}, \mathbf{q}| U_{0 0}(0) | \mathbf{0}, \mathbf{0} \rangle$. The shaded diagrams are not included in the operator $\mathcal{T}$ of Ref.~\cite{braaten2002diluteBEC}. The vertices represent the two-body interaction. }
    \label{fig:Feynman}
\end{figure}

Braaten et al. \cite{braaten2002diluteBEC} have calculated a three-body elastic scattering amplitude that is closely related to the amplitude $\langle \mathbf{0}, \mathbf{q} | U_{00}(0) | \mathbf{0}, \mathbf{0} \rangle$ that we calculate. The main difference is that the three-body operator $T$ as defined in Ref.~\cite{braaten2002diluteBEC} (which we denote by $\mathcal{T}$ in the following as we already reserved $T$ for the two-body transition operator) includes less scattering events than the AGS operator $U_{00}$. First of all, scattering processes in which the third particle spectates are not included in $\mathcal{T}$. Secondly, two-body scattering events that occur before the third particle participates in the scattering process are incorporated in so-called dimer propagators \cite{braaten2002diluteBEC, braaten2006universality} and are thus not included in $\mathcal{T}$. The same applies to two-body scattering events that occur after the third particle has moved away from the two-particle subsystem. For clarity, we show some Feynman diagrams for three-body elastic scattering in \refFigure{fig:Feynman}. This figure indicates which scattering events are not included in $\mathcal{T}$.

Consequently, some terms on the right-hand-side of \refEquation{eq:TG0PT(1+P)_matrix} are not included in the calculation of $\langle \mathbf{0}, \mathbf{q} | \mathcal{T}(0) | \mathbf{0}, \mathbf{0} \rangle$, namely
$
- 3 \sqrt{3}/(4 \pi^4) a^3/(m \hbar^3 q_{\alpha})
$
 and
\begin{equation}
\begin{aligned}
-3 m & \langle \mathbf{0}| T(0) | \mathbf{0} \rangle  \frac{\partial^2 \langle \mathbf{0}| T\left(-\frac{3}{4 m} q^2\right) | \mathbf{0} \rangle}{\partial q^2}\Bigg|_{q=0} 
\\
&= -\frac{9}{8 \pi^4} \frac{a^4}{m \hbar^4} + O(a^3)
\end{aligned}
\end{equation}
where terms that grow less fast than $a^4$ are indicated by $O(a^3)$. A similar analysis reveals that the right-hand-side of \refEquation{eq:TG0PTG0PT(1+P)_matrix} includes the term
$
\sqrt{3}/(2 \pi^3) a^4/(m \hbar^4)
$
which is absent in $\langle \mathbf{0}, \mathbf{q} | \mathcal{T}(0) | \mathbf{0}, \mathbf{0} \rangle$.
All other scattering events that are incorporated in $\langle \mathbf{0}, \mathbf{q} | U_{0 0}(0) | \mathbf{0}, \mathbf{0} \rangle$, but not in $\langle \mathbf{0}, \mathbf{q} | \mathcal{T}(0) | \mathbf{0}, \mathbf{0} \rangle$ vanish in the limit $q \to 0$ or grow less fast than $a^4$. So from this analysis combined with \del{Eq.~(3) of the main text}, the connection between the scattering hypervolume $D$ and the quantity $A(a \Lambda_*)$ as defined by Ref.~\cite{braaten2002diluteBEC} can be made resulting in 
\begin{equation}
\frac{1}{C}\Bigg[\frac{1}{a^4}D
  -
(2 \pi)^6 \Bigg(\frac{\sqrt{3}}{2 \pi^3} -\frac{9}{8 \pi^4}\Bigg) \Bigg]
\underset{|a| \to \infty}{=}  A(a \Lambda_*).
\end{equation}
The three-body parameter $\Lambda_*$ depends on the considered two-body potential and is related to $a_-$ and $a_+$ \cite{braaten2006universality}.

\section{Calculation of the three-body elastic scattering amplitude}

Here we comment on the approach that we use to calculate the three-body elastic scattering amplitude $\langle \mathbf{p}, \mathbf{q} | U_{00}(0) | \mathbf{0}, \mathbf{0} \rangle$.
In fact, we use the same method as presented in Ref.~\cite{mestrom2019squarewell} to solve \refEquation{eq:Ubreve}
for the matrix elements ${}_{\alpha} \langle \mathbf{p}, \mathbf{q} | \breve{U}_{\alpha 0}(0) | \mathbf{0}, \mathbf{0} \rangle$ from which the elastic scattering amplitude is determined via \refEquation{eq:U00_vs_Ubreve_matrix}.
So we write down the integral equations for the matrix elements ${}_{\alpha} \langle \mathbf{p}, \mathbf{q} | \breve{U}_{\alpha 0}(0) | \mathbf{0}, \mathbf{0} \rangle$, and expand them in partial waves and in form factors that are determined by the Weinberg expansion \cite{weinberg1963expansion, mestrom2019squarewell}. The resulting one-dimensional integral equation can be solved as a matrix equation by discretizing the magnitude of the momentum $q$.
The amplitude ${}_{\alpha} \langle \mathbf{p}, \mathbf{q} | \breve{U}_{\alpha 0}(0) | \mathbf{0}, \mathbf{0} \rangle$ diverges as $\delta(q)$, $q^{-2}$, $q^{-1}$ and $\text{ln}(q)$ in the limit $q\to 0$. We deal with these singularaties by subtracting them from the matrix elements and solve the integral equations for the remaining part. To clarify this approach, we define the ket state $\breve{U}_{\alpha 0}^{(\text{div})} \lvert \mathbf{0}, \mathbf{0}\rangle$ whose momentum-space representation is given by
\begin{equation}
\begin{aligned}
{}_{\alpha}\langle \mathbf{p}, &\mathbf{q} | \breve{U}_{\alpha 0}^{(\text{div})} | \mathbf{0}, \mathbf{0}  \rangle
\equiv 
3 \, \delta( \mathbf{q} )
\langle \mathbf{p} | T(0) | \mathbf{0} \rangle
\\
&
-\frac{3}{2 \pi^4} \frac{a^2}{m \hbar^2} \frac{1}{q^2}
\\
&
+ \frac{1}{4 \pi^4} (4 \pi - 3 \sqrt{3})
\frac{a^3}{m \hbar^3} \frac{1}{q}
\\
&
+
\frac{1}{\pi^5} (4 \pi - 3 \sqrt{3}) \frac{a^4}{m \hbar^4} \text{ln}\Bigg(\frac{q |a|}{\hbar}\Bigg).
\end{aligned}
\end{equation}
If we then define the state $\breve{U}_{\alpha 0}^{(\text{nondiv})} \lvert \mathbf{0}, \mathbf{0}\rangle$ by
\begin{equation}
\breve{U}_{\alpha 0}^{(\text{nondiv})} \lvert \mathbf{0}, \mathbf{0}\rangle
\equiv
 \breve{U}_{\alpha 0}(0) \lvert \mathbf{0}, \mathbf{0}\rangle
- \breve{U}_{\alpha 0}^{(\text{div})} \lvert \mathbf{0}, \mathbf{0}\rangle,
\end{equation}
we can derive from \refEquation{eq:Ubreve} the following inhomogeneous equation for $\breve{U}_{\alpha 0}^{(\text{nondiv})} \lvert \mathbf{0}, \mathbf{0}\rangle$:
\begin{equation}\label{eq:Ubreve_nondiv} 
\begin{aligned}
\breve{U}_{\alpha 0}^{(\text{nondiv})} \lvert \mathbf{0}, \mathbf{0}\rangle &= T_{\alpha}(0)(1+P) \lvert  \mathbf{0}, \mathbf{0}\rangle - \breve{U}_{\alpha 0}^{(\text{div})} \lvert \mathbf{0}, \mathbf{0} \rangle
\\
&
+ T_{\alpha}(0) G_0(0) P \breve{U}_{\alpha 0}^{(\text{div})} \lvert \mathbf{0}, \mathbf{0}\rangle
\\
&
+ T_{\alpha}(0) G_0(0) P \breve{U}_{\alpha 0}^{(\text{nondiv})} \lvert \mathbf{0}, \mathbf{0} \rangle.
\end{aligned}
\end{equation}
The corresponding momentum-space representation ${}_{\alpha} \langle \mathbf{0}, \mathbf{q} | \breve{U}_{\alpha 0}^{(\text{nondiv})} | \mathbf{0}, \mathbf{0}\rangle$ is free of divergences as $q \to 0$ and can be calculated numerically from \refEquation{eq:Ubreve_nondiv} as described above.

The number of partial-wave components that we need to take into account increases as the potential depth increases \cite{mestrom2019squarewell}. For three-body calculations near the first two-body $s$-wave potential resonance of the square-well potential, it suffices to take $l = 0$ and $l = 2$, and to neglect the higher partial-wave components for a relative uncertainty of $10^{-3}$. Near the second potential potential resonance, we take $l = 0$, 2, 4, 6, and 8 for the same precision, whereas for small scattering lengths in between the second and third potential resonance we also take $l = 10$ into account. The total number of Weinberg expansion terms that we take into account varies from 13 near the first potential resonance to 59 in between the second and third potential resonance. 

\begin{figure}[hbtp]
    \centering
    \includegraphics[width=3.4in]{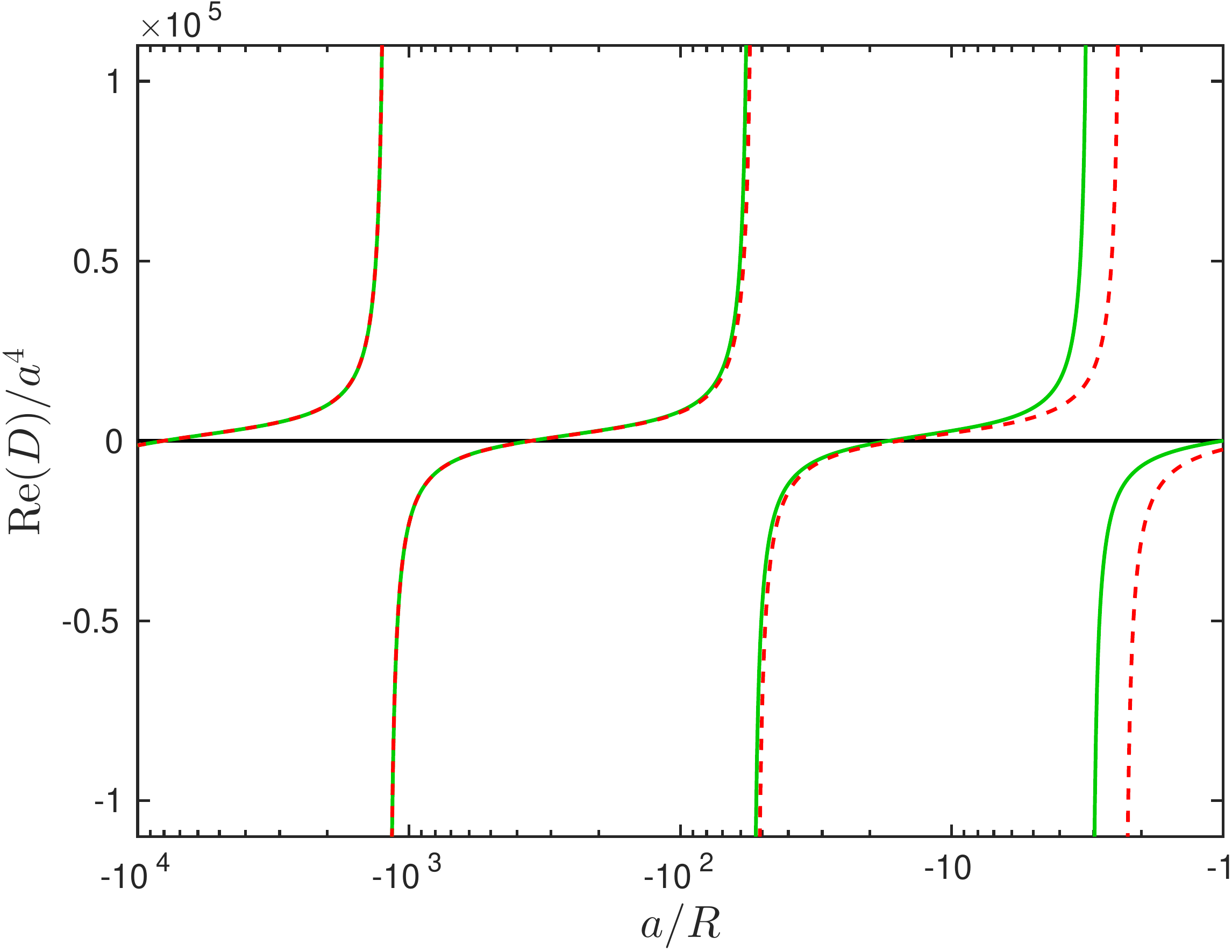}
    \caption{Three-body scattering hypervolume $D$ (green solid line) near the first potential resonance of the square-well potential for $a<0$. The dashed curve gives the analytic zero-range results given by \del{Eq.~(8) of the main text} where we set $a_-/R = -1205$, $b_- = 3.153$, $c_- = 1.140$ and $\eta = 0$.}
    \label{fig:SqW_Res1_Dhyp_a0neg_-1_-1e4}
\end{figure}
% Figure made with the code: Plot_and_fit_Elastic_K3_List_v9_SqW_Res1_a0neg_v2_No_fit.m

\begin{figure}[hbtp]
    \centering
    \includegraphics[width=3.4in]{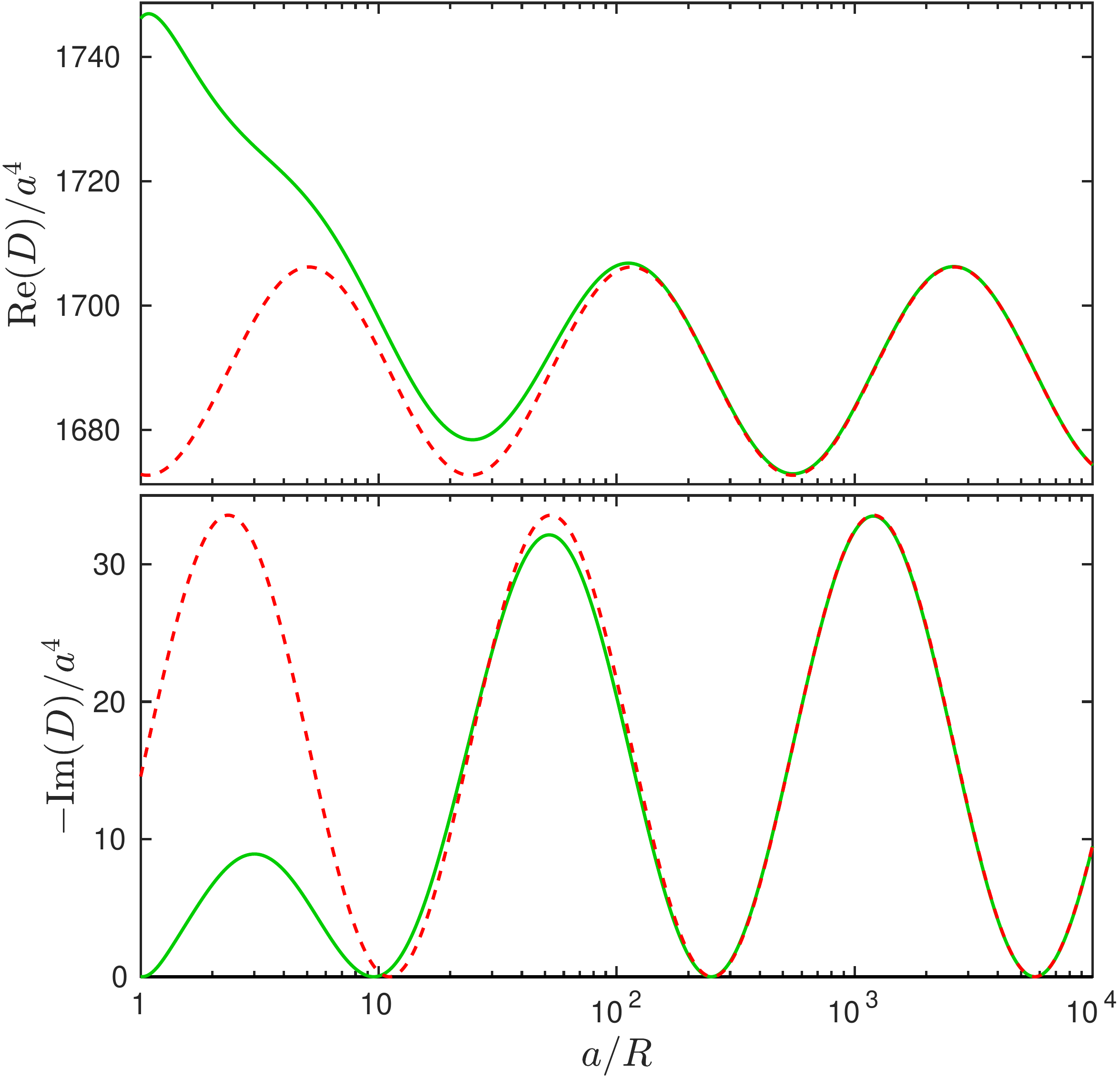}
    \caption{Three-body scattering hypervolume $D$ (green solid line) near the first potential resonance of the square-well potential for $a>0$. The dashed curves give the analytic zero-range results given by \del{Eq.~(9) and (11) of the main text} where we set $a_+/R = 5720$, $b_+ = 0.0226$, $c_+ = 1.1288$, $C_+ = 67.118$ and $\eta = 0$.} 
    \label{fig:SqW_Res1_Dhyp_a0pos_1_1e4}
\end{figure}
% Figure made with the code: Plot_and_fit_Elastic_K3_List_v9_SqW_Res1_a0pos_v2_No_fit_Im.m

\begin{table*}[htb!]
  \centering
  \caption{Values of the characteristic parameters of the three-body scattering hypervolume $D$ (real and imaginary part) corresponding to the $N$th potential resonance of the square-well potential. The three-body parameters $a_{-,n}$ and $a_{+,n}$ locate the trimer resonances corresponding to the $(n+1)$th Efimov state. They are obtained by fitting the data close to the resonance position with \del{Eq.~(8) of the main text}
  %\cref{eq:Dhyp-_universal_with_eta} 
  for $a_{-,n}$ and 
  \del{Eq.~(11) of the main text}
  %\cref{eq:K3_a0pos_universal_eta} 
  for $a_{+,n}$. We only consider the three-body parameters for $|a|/R<10^4$. Consequently, we have not determined $a_{+,3}$ for $N  = 2$. The loss parameter $\eta$ is determined by fitting the numerical data in the strongly interacting regime ($10^3 < |a|/R <10^4$).}
  \label{tab:SqW_3BP_3body_elastic}
  \begin{ruledtabular}
    \begin{tabular}{ccccccccc}
    \toprule
    $N$ & $a_{-,0}/R$ & $a_{-,1}/R$ & $a_{-,2}/R$ & $a_{+,0}/R$ & $a_{+,1}/R$ & $a_{+,2}/R$ & $a_{+,3}/R$ & $\eta$ \\
    \hline
    1 & -3.087(1) & -54.90(1) & -1205(1) & 1.015(1) & 9.530(1) & 249.7(1) & 5720(1) & 0 \\
    2 & -17.43(1) & -372.1(1) & -8396(1) & 1.21(5) & 75.7(1) & 1759(5) & - & 0.068(1)\\
    \bottomrule
    \end{tabular}\\
    \end{ruledtabular}
\end{table*}
% See my analysis of Tuesday 29 January 2019, schrift 20, PhD jaar 2.

\section{Additional results for the first and second potential resonance}

We have also calculated the three-body scattering hypervolume $D$ in the strongly interacting regime near the first $s$-wave potential resonance corresponding to the square-well interaction. These results are presented in \refFigure{fig:SqW_Res1_Dhyp_a0neg_-1_-1e4} for $a<0$ and in \refFigure{fig:SqW_Res1_Dhyp_a0pos_1_1e4} for $a>0$, in which they are compared to the analytic zero-range results. An overview of the three-body parameters $a_-$ and $a_+$ is given in \refTable{tab:SqW_3BP_3body_elastic}.
Our result for the loss parameter $\eta$ in \refTable{tab:SqW_3BP_3body_elastic} is the same for large positive and negative scattering lengths near the second potential resonance and is consistent with the results of Ref.~\cite{mestrom2019squarewell} from which it can be infered that $\eta$ is expected to be in between $0.06$ and $0.08$ for $|a|/R \gg 1$. The values for $a_{-,0}$ stated in \refTable{tab:SqW_3BP_3body_elastic} also agree with those presented in Ref.~\cite{mestrom2019squarewell}.

\begin{figure}[hb!]
    \centering
    \includegraphics[width=3.4in]{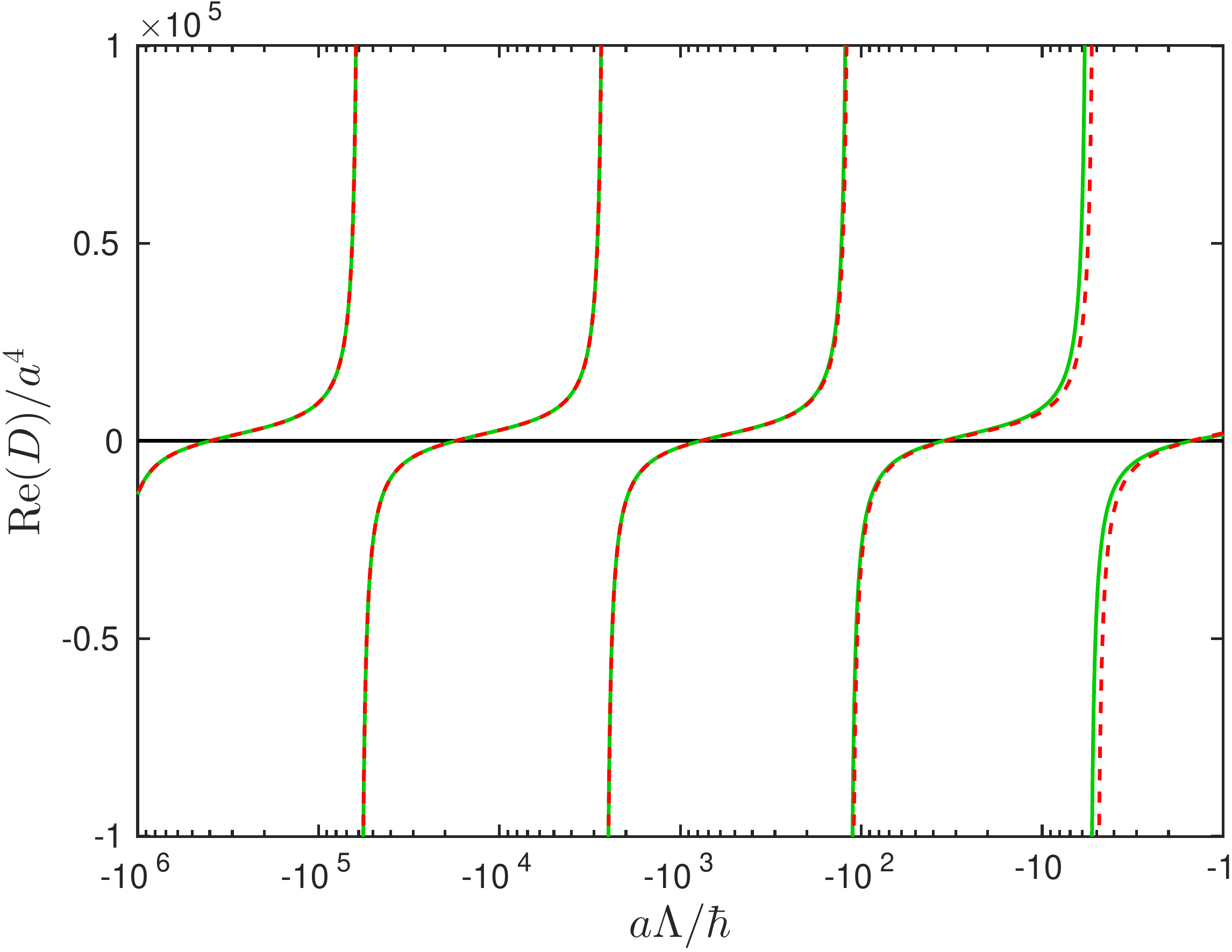}
    \caption{Three-body scattering hypervolume $D$ (green solid line) near the potential resonance of the contact interaction with momentum cutoff $\Lambda$ for $a<0$. The dashed curve gives the analytic zero-range results given by \del{Eq.~(8) of the main text} where we set $a_- \Lambda/\hbar = -5.926 \cdot 10^4$, $b_- = 3.153$, $c_- = 1.140$ and $\eta = 0$.}
    \label{fig:sepCut_Dhyp_a0neg_-1_-1e6}
\end{figure}
% Figure made with the code Plot_and_fit_Elastic_K3_List_v19_sepCut_a0neg_No_fit_plotted.m

\begin{figure}[hb!]
    \centering
    \includegraphics[width=3.4in]{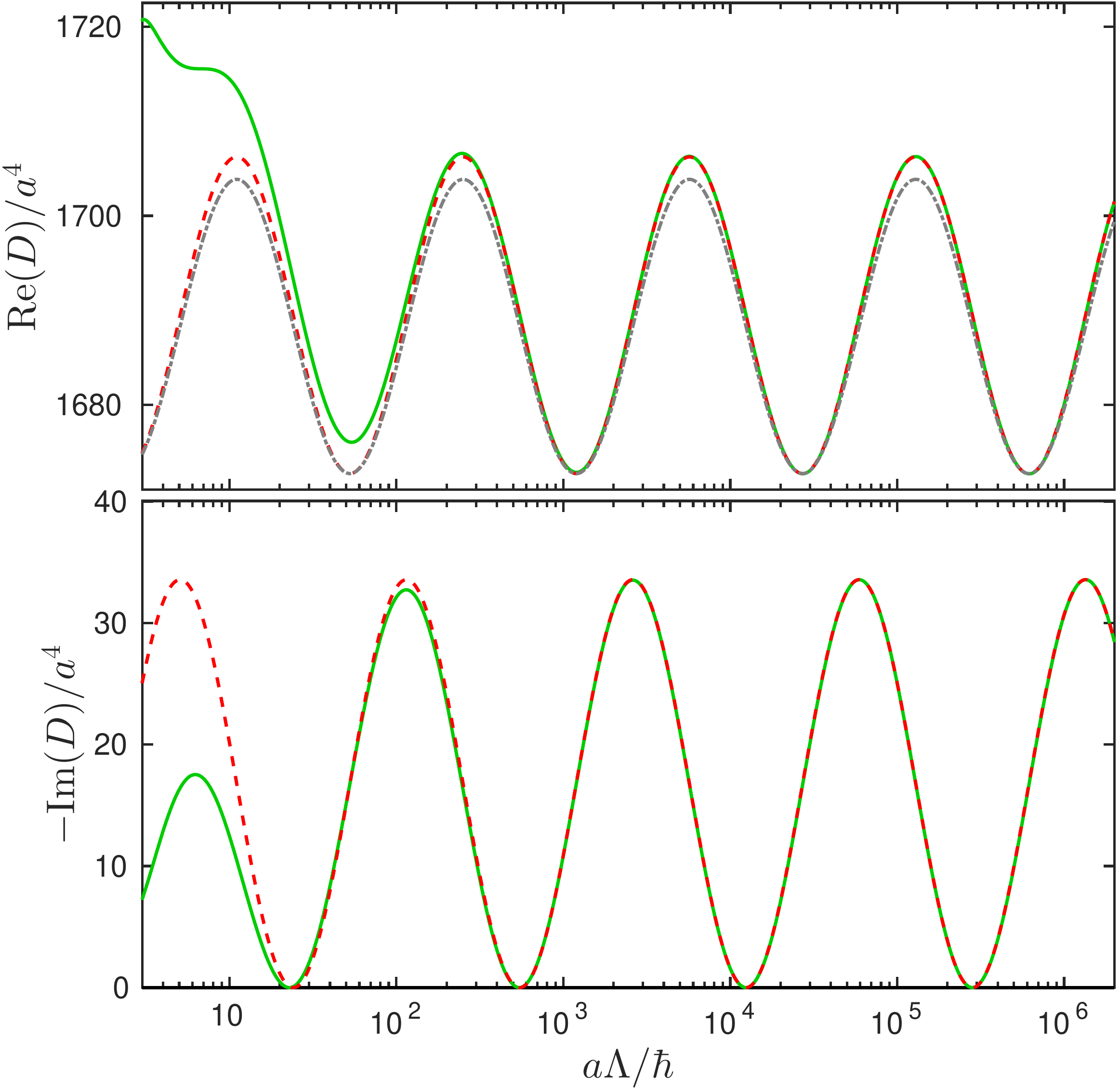}
    \caption{Three-body scattering hypervolume $D$ (green solid line) near the potential resonance of the contact interaction with momentum cutoff $\Lambda$ for $a>0$. The red dashed curves give the analytic zero-range results given by \del{Eq.~(9) and (11) of the main text} where we set $a_+ \Lambda/\hbar = 2.82 \cdot 10^5$, $b_+ = 0.0226$, $c_+ = 1.1288$, $C_+ = 67.118$ and $\eta = 0$. The gray dash-dotted curve displays the analytic zero-range result given by \del{Eq.~(9) of the main text} with the same parameters except for $b_+ = 0.021$ which is the value reported in Ref.~\cite{braaten2002diluteBEC}. } 
    \label{fig:sepCut_Dhyp_a0pos_1_2e6}
\end{figure}
% Figure made with the code Plot_and_fit_Elastic_K3_List_v20_sepCut_a0pos_No_fit_plotted_BH.m

\section{Universal parameters}

The universal coefficients appearing in the analytic zero-range results \del{(see Eq.~(8)$-$(11) in the main text)} can be most easily determined from the contact interaction itself. Since we work in momentum space, we add a momentum cutoff $\Lambda$ to the contact interaction, i.e.,
\begin{equation} \label{eq:V_sepCut}
V = - \zeta \lvert g \rangle \langle g \rvert,
\end{equation}
where
\begin{equation}
\langle \mathbf{p} | g \rangle = 
\begin{cases}1,&\mbox{$0\leq p \leq \Lambda$},\\0,&\mbox{$p> \Lambda$}.\end{cases}
\end{equation}
The scattering length can be changed by tuning the interaction strength $\zeta$. The three-body scattering hypervolume $D$ is calculated according to the same procedure as presented in the main text. Since the contact interaction is separable, there is only one potential resonance and the loss parameter $\eta$ that appears in the universal equations \del{(see Eq.~(8)$-$(11) in the main text)} is zero.

Our results for the contact interaction with a momentum cutoff are shown in \refFigure{fig:sepCut_Dhyp_a0neg_-1_-1e6}  for $a<0$ and in \refFigure{fig:sepCut_Dhyp_a0pos_1_2e6} for $a>0$. We determine the universal coefficients $C_+$, $b_{\pm}$ and $c_{\pm}$ from the characteristic points of these curves. First of all, the coefficient $C_+$ can be determined from the local maxima of $-\text{Im}(D/a^4)$ as presented in \refTable{tab:sepCut_3body_elastic_C+}. Our results show that $C_{+,n}$ converges to \del{$C_+ = 67.118(5)$} for $a \to \infty$.

Similarly, $b_+$ and $c_+$ can be determined from the local maxima and minima of Re($D/a^4$). The results of our analysis can be found in \refTable{tab:sepCut_3body_elastic_b+_c+} from which we conclude that \del{$b_+ = 0.0226(5)$ and $c_+ = 1.1288(5)$}.

For negative scattering lengths, $D$ is real and diverges at scattering lengths equal to the three-body parameters $a_{-,n}$. The behavior of $D$ is given by \del{Eq.~(8) of the main text}. By multiplying $D/(C a^4)$ with $\sin\Big( s_0 \ \text{ln}(a/a_{-,n}) \Big)$, we remove the singularity in $D$ at $a = a_{-,n}$. The resulting quantity behaves at scattering lengths near $a_{-,n}$ as
\begin{equation}
\begin{aligned}
D/(C a^4) &\sin\Big( s_0 \ \text{ln}(a/a_{-,n}) \Big)=
\sqrt{b_{-,n}^2 + c_{-,n}^2} 
\\
&
\cos\Big( s_0 \ \text{ln}(a/a_{-,n}) - \delta_{-,n}\Big)
\end{aligned}
\end{equation}
where
\begin{equation}
\cos(\delta_{-,n}) = \frac{b_{-,n}}{\sqrt{b_{-,n}^2 + c_{-,n}^2}} 
\end{equation}
or
\begin{equation}
\sin(\delta_{-,n}) = \frac{c_{-,n}}{\sqrt{b_{-,n}^2 + c_{-,n}^2}}. 
\end{equation}
We determine the amplitude $\sqrt{b_{-,n}^2 + c_{-,n}^2}$ and phase shift $\delta_{-,n}$ from our numerical data of $D$. The resulting values of $b_{-,n}$ and $c_{-,n}$ can be found in \refTable{tab:sepCut_3body_elastic_b-_c-}. The coefficients $c_{-,n}$ converge faster than $b_{-,n}$ and we conclude that \del{$b_- = 3.153(5)$ and $c_- = 1.140(2)$}.

\begin{table}[htb!]
  \centering
  \caption{Parameters corresponding to the local maxima of $-2 \cdot \text{Im}(D/a^4)$ calculated for the potential given by \refEquation{eq:V_sepCut}. The local maxima are indicated by $C_{+,n}$ and the scattering lengths at which these maxima occur are indicated by $a_{\text{Im},n}$.
  %\ins{13 March 2019: I checked the uncertainties amaxIm}
  }
  \label{tab:sepCut_3body_elastic_C+}
    \begin{ruledtabular}
    \begin{tabular}{ccc}
    \toprule
    $n$ & $a_{\text{Im},n} \Lambda/\hbar$ & $C_{+,n}$ \\
    \hline
    2 & $1.145(5)\cdot 10^2$ & 65.453(5)  \\
    3 & $2.61(1)\cdot 10^3$ & 67.045(5) \\
    4 & $5.93(1)\cdot 10^4$ & 67.114(5) \\
    5 & $1.344(3)\cdot 10^6$ & 67.118(5) \\
    \bottomrule
    \end{tabular}\\
    \end{ruledtabular}
\end{table}

\begin{table}[htb!]
  \centering
  \caption{Parameters corresponding to the local maxima and minima of $\text{Re}(D/a^4)/C$ calculated for the potential given by \refEquation{eq:V_sepCut}. The local minima are indicated by $c_{+,n}$ and are located at $a = a_{\text{min},n}$, whereas $b_{+,n}$ is determined from the difference between the local maximum at $a = a_{\text{max},n}$ and minimum at $a = a_{\text{min},n}$. 
  %The scattering lengths at which these maxima occur are indicated by $a_{\text{max},n}$.  
  %\ins{12 March and 13 March 2019: I checked the uncertainties amin and amax}
  }
  \label{tab:sepCut_3body_elastic_b+_c+}
   \begin{ruledtabular}
    \begin{tabular}{ccccc}
    \toprule
    $n$ & $a_{\text{max},n}  \Lambda/\hbar$ & $a_{\text{min},n}  \Lambda/\hbar$ & $b_{+,n}$ & $c_{+,n}$ \\
    \hline
    1 & - & $5.39(2) \cdot 10^1$ & - & 1.1310(5) \\
    2 & $2.47(1) \cdot 10^2$ & $1.198(5) \cdot 10^3$ & 0.02277(50) & 1.1289(5) \\
    3 & $5.68(3) \cdot 10^3$ & $2.72(1) \cdot 10^4$ & 0.02265(50) & 1.1288(5) \\
    4 & $1.29(3) \cdot 10^5$ & $6.17(5) \cdot 10^5$ & 0.02265(50) & 1.1288(5) \\
    \bottomrule
    \end{tabular}\\
    \end{ruledtabular}
\end{table}
% See my code Plot_and_fit_Elastic_K3_List_v14_sepCut_a0pos.m. Analyzed on Wednesday 6 March 2019, schrift 21, PhD jaar 2.
% Better: See my code Plot_and_fit_Elastic_K3_List_v17_sepCut_a0pos.m.

\begin{table}[htb!]
  \centering
  \caption{Values of $b_{-,n}$ and $c_{-,n}$ that are determined from the amplitude and phase shift of the oscillatory function $D/(C a^4) \sin\Big( s_0 \ \text{ln}(a/a_{-,n}) \Big)$ for the potential given by \refEquation{eq:V_sepCut}. The amplitude $\sqrt{b_{-,n}^2 + c_{-,n}^2}$ and phase shift $\delta_{-,n}$ are determined near the $(n+1)$th Efimov resonance that occurs at $a = a_{-,n}$.}
  \label{tab:sepCut_3body_elastic_b-_c-}
    \begin{ruledtabular}
    \begin{tabular}{cccccc}
    \toprule
    $n$ & $a_{-,n} \Lambda/\hbar$ & $\delta_{-,n}$ & $\sqrt{b_{-,n}^2 + c_{-,n}^2}$ & $b_{-,n}$ & $c_{-,n}$ \\
    \hline
   $1$ & $-1.169(1)\cdot 10^2$ & 0.3491(10) & 3.3175(5) & 3.117(2) & 1.135(3) \\
   $2$ & $-2.614(1)\cdot 10^3$ & 0.3473(10) & 3.3497(5) & 3.150(2) & 1.140(3) \\
   $3$ & $-5.926(1)\cdot 10^4$ & 0.3470(10) & 3.3525(5) & 3.153(2) & 1.140(3) \\
    \bottomrule
    \end{tabular}\\
    \end{ruledtabular}
\end{table}
% See my code Plot_and_fit_Elastic_K3_List_v19_sepCut_a0neg.m. Analyzed on Thursday 7 March 2019 AND Tuesday 12 March 2019, schrift 21, PhD jaar 2.

\newpage
\section{Analysis of the eigenvalues of the kernel}

\begin{figure}[hbtp]
    \centering
    \includegraphics[width=3.4in]{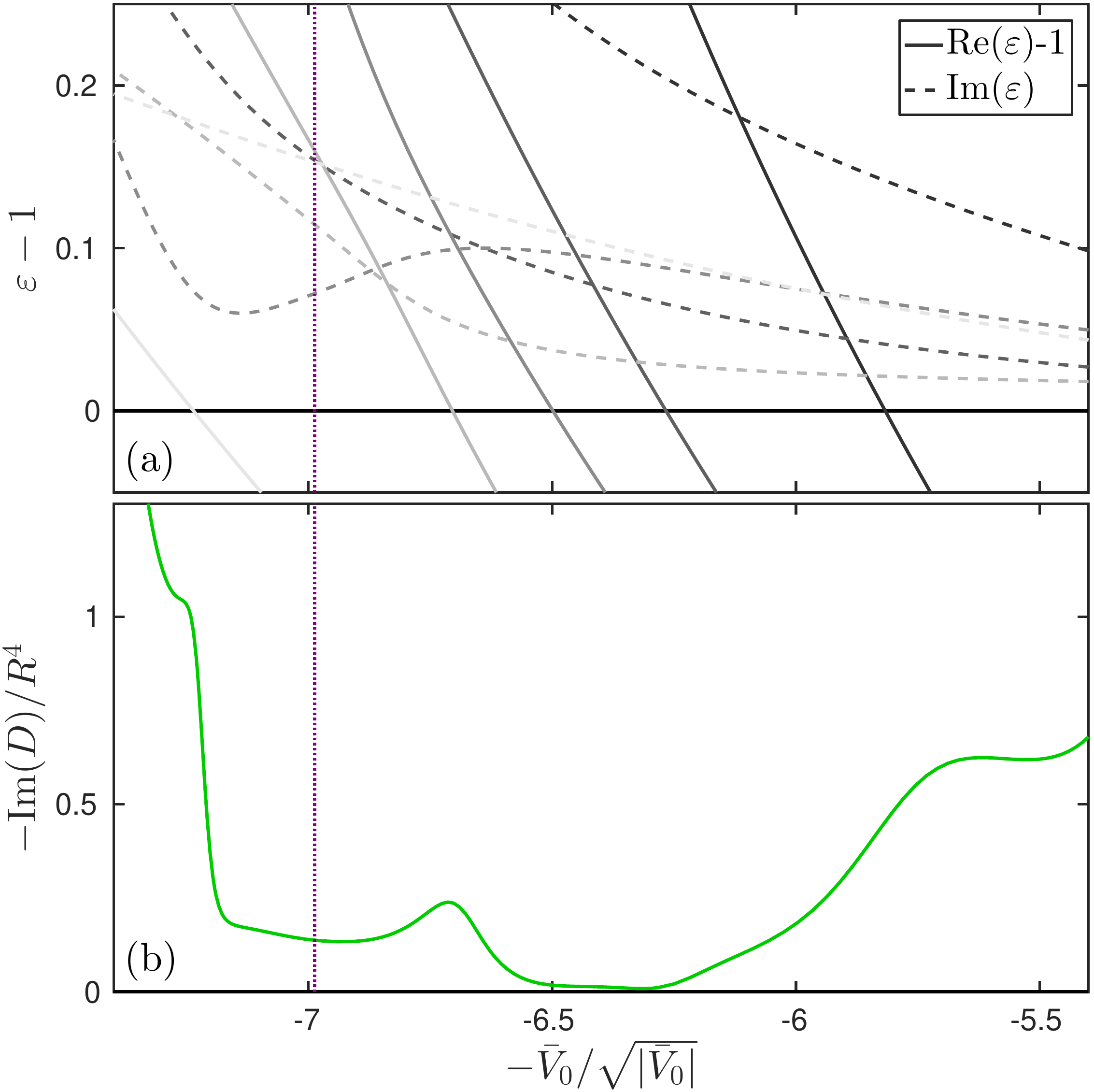}
    \caption{Eigenvalues $\varepsilon$ (a) of the kernel of the three-body integral equation corresponding to resonances in the imaginary part of the three-body scattering hypervolume (b). These quantities are shown as a function of the dimensionless interaction strength $\bar{V}_0$ of the square-well potential. The vertical purple dotted line indicates the interaction strength at which the first $g$-wave dimer state becomes bound.}
    \label{fig:SqW_q0_5d4_7d4_ImD_eigval}
\end{figure}

As discussed in Ref.~\cite{mestrom2019squarewell}, the eigenvalues of the kernel of the three-body integral equation indicate the position of three-body resonances. The real part of the eigenvalue $\varepsilon$ that corresponds to the resonance passes one near the resonance position. Figure~\ref{fig:SqW_q0_5d4_7d4_ImD_eigval} shows that the real part of $\varepsilon$ indeed passes one at the trimer resonance positions $\sqrt{\bar{V_0}} = 5.8$, $6.7$, and $7.2$. This proofs that the observed resonances are caused by three-body quasibound states. The eigenvalues whose real part passes one at $\sqrt{\bar{V_0}} = 6.3$ and $6.5$ might also correspond to  metastable three-body states although there is no clear resonance structure in $\text{Im}(D)$.

% Figure made with the code: Plot_and_fit_Elastic_K3_List_v29_SqW_Res_2and3_ImD_eig.m

\end{document}